\documentclass[%
 reprint,
showpacs,preprintnumbers,
 amsmath,amssymb, aps, prb,
]{revtex4-1}
\usepackage{graphicx}
\usepackage{amsmath}
\usepackage{amsfonts}
\usepackage{amssymb}
\usepackage{amscd}
\usepackage{cancel}
\usepackage{ntheorem}
\usepackage[american]{babel}
\usepackage[T1]{fontenc}
\usepackage{textcomp}
\usepackage[dvips]{rotating}
\usepackage{fancybox}
\usepackage{eurosym}
\usepackage[utf8]{inputenc}
\usepackage{epsfig}
\usepackage[pdftex]{color}
\usepackage{graphicx}
\usepackage{palatino}
\usepackage{array}

\renewcommand{\v}[1]{\ensuremath{\mathbf{#1}}} % for vectors
 
\newcommand{\avg}[1]{\left< #1 \right>} % for average
\let\baraccent=\= % rename builtin command \= to \baraccent
\renewcommand{\=}[1]{\stackrel{#1}{=}} % for putting numbers above 

\renewcommand{\b}[1]{{\bf #1}}
\renewcommand{\i}{\mathrm{i}}
\def\beq#1\eeq{\begin{equation}\begin{split}#1\end{split}\end{equation}}

\newcommand{\bmat}{\begin{pmatrix}}
\newcommand{\emat}{\end{pmatrix}}
\newcommand{\bdet}{\begin{dmatrix}}
\newcommand{\edet}{\end{dmatrix}}

\newcommand{\Lra}{\Leftrightarrow}

\newcommand{\tn}[1]{\textnormal{#1}}

\newcommand{\Comm}[1]{\Bigl[\,#1\,\Bigr]}
\makeatletter
\newcommand{\showfont}{encoding: \f@encoding{},
  family: \f@family{},
  series: \f@series{},
  shape: \f@shape{},
  size: \f@size{}
}
\newcommand{\iffont}[3]{\ifthenelse{\equal{\f@family}{#1}}{#2}{#3}}
\makeatother

\begin{document}

%\preprint{}

\title{Hubbard nanoclusters far from equilibrium}



\author{Sebastian Hermanns}
\email[]{hermanns@theo-physik.uni-kiel.de}
\author{Niclas Schl\"unzen}
\author{Michael Bonitz}
%\thanks{}
\affiliation{Institut f\"ur Theoretische Physik und Astrophysik, Christian-Albrechts-Universit\"at zu Kiel, Leibnizstra\ss{}e 15, D-24098 Kiel, Germany
}

\newcommand{\eq}[1]{Eq.~(\ref{#1})}
\newcommand{\nn}{\nonumber}
\newcommand{\e}[1]{\mathrm{e}^{#1}}
\newcommand{\op}[1]{\hat{#1}}
\newcommand{\eqsand}[2]{Eqs.~(\ref{#1}) and~(\ref{#2})}
\newcommand{\cc}{{\cal C}}
\newcommand{\tc}{T_{\cal C}}
\newcommand{\dc}{\delta_{\cal C}}
\newcommand{\thc}{\theta_{\cal C}}
\newcommand{\intc}[1]{\int_{\cal C}\mathrm{d}{#1}\;}
\newcommand{\intlim}[3]{\int_{#1}^{#2}\mathrm{d}{#3}\;}
\newcommand{\mret}{{\mathrm{ret}}}
\newcommand{\madv}{{\mathrm{adv}}}

\date{\today}

\begin{abstract}
The Hubbard model is a prototype for strongly correlated many-particle systems, including electrons in condensed matter 
and molecules, as well as for fermions or bosons in optical lattices. While the equilibrium properties of these systems 
have been studied in detail, the nonequilibrium dynamics following a strong non-perturbative excitation only recently 
came into the focus of experiments and theory. It is of particular interest how the dynamics depend on the coupling 
strength and on the particle number and whether there exist universal features in the time evolution.
Here, we present results for the dynamics of finite Hubbard clusters based on a selfconsistent nonequilibrium Green 
functions (NEGF) approach invoking the generalized Kadanoff--Baym ansatz (GKBA). We discuss the conserving properties of 
the GKBA with Hartree--Fock propagators in detail and present a generalized form of the energy conservation criterion of 
Baym and Kadanoff for NEGF.
Furthermore, we demonstrate that the HF-GKBA cures some artifacts of prior two-time NEGF 
simulations. Besides, this approach substantially speeds up the numerical calculations and thus presents the capability to 
study comparatively large systems and to extend the analysis to long times 
allowing for an accurate computation of the excitation spectrum via time propagation. Our data obtained within the second Born approximation 
compares favorably with exact diagonalization results (available for up to 13 particles) and are expected to have predictive capability for 
substantially larger systems in the weak coupling limit.
\end{abstract}


\maketitle

% --------------------------
% --- Introduction ---------
\section{Introduction}\label{s:intro}
Strongly correlated quantum systems and materials, e.g.~[\onlinecite{pavarini11}], are of rapidly growing relevance 
in many fields of physics and chemistry. Especially the out-of-equilibrium dynamics are of great current interest in 
solid-state, atomic and molecular physics, in nanoelectronics, quantum transport etc. In all these fields, the 
availability of intense and coherent radiation, combined with ultra-short laser pulses, has triggered many key 
experiments that allow one to investigate matter under extreme nonequilibrium conditions where strong correlations and 
nonlinear effects occur simultaneously [\onlinecite{balzer_jpcs13}]. Examples are the photoionization of multi-electron atoms 
and molecules, e.g.~[\onlinecite{becker12}, \onlinecite{schuette_prl12}] and references therein. Another example are 
the many-body dynamics of particles in lattice systems in condensed matter, 
e.g.~[\onlinecite{kehrein_njp10}, \onlinecite{eckstein_prl09}, \onlinecite{eckstein_prb10}] and
optical lattices~[\onlinecite{bloch08}] following a rapid quench of the interaction strength.

From the theoretical point of view, such systems pose particular challenges since quantum, spin and strong 
correlation effects have to be treated selfconsistently under situations far from the ground state or from 
thermodynamic equilibrium. Here, remarkable progress has been achieved recently using ab initio methods such as exact 
diagonalization, density matrix renormalization group approaches, nonequilibrium dynamical mean field theory 
[\onlinecite{eckstein_prl09}], iterative path integral techniques [\onlinecite{thorwart_13}] and others. The common problem of these approaches is an exponential scaling with the particle number and restrictions with respect to the duration of the time propagation.
%(\textcolor{red}{TD-DMRG kann auch lange Zeiten}).

For this reason, alternative approaches that are based on statistical methods are of high interest. This includes 
density operator methods, e.g. [\onlinecite{book_bonitz_qkt,akbari_12,hermanns_jpcs13}], 
nonequilibrium Green function (NEGF) techniques and a recently developed stochastic mean field approach [\onlinecite{ayik09,lacroix14}]. 
Here, we focus on the NEGF method which, during the past 15 years, has been successfully applied to 
a variety of many-body systems in nonequilibrium, including the optical excitation of electron-hole plasmas in 
semiconductors~[\onlinecite{kwong98}, \onlinecite{kwong00}], nuclear collisions~[\onlinecite{rios11}], dynamics of laser plasmas 
[\onlinecite{bonitz_cpp99}, \onlinecite{haberland_pre01}] and the problem of baryogenesis in cosmology~[\onlinecite{garny11}]. More recently, NEGF 
methods have also been used to describe finite spatially inhomogeneous systems, including 
the carrier dynamics and carrier-phonon interaction in quantum dots and quantum 
wells~[\onlinecite{gartner06,lorke06,bonitz_prb07,balzer_prb09}], molecular transport in contact with leads, 
e.g.~[\onlinecite{uimonen11, khosravi12, stefanucci_gkba}], or small atoms, 
e.g.~[\onlinecite{dahlen07}, \onlinecite{balzer_pra10}, \onlinecite{balzer_pra10_2}]. For a recent overview on NEGF 
applications to inhomogeneous systems, see Refs.~[\onlinecite{balzer_lnp13}, \onlinecite{stefanucci_book_13}].

Applications of NEGF methods to small Hubbard clusters have been presented not long ago 
[\onlinecite{puigvonfriesen09,puigvonfriesen10}] and showed the great potential of this method. The physical features that could be explored include the relaxation dynamics, the excitation spectrum and, in particular, the relevance of double 
excitations [\onlinecite{balzer12_epl,sakkinen12}]. At the same time, NEGF simulations exhibited fundamental problems: The first is of conceptual nature and is related to unphysical damping effects in case of strong excitation which are absent in exact calculations [\onlinecite{puigvonfriesen09}]. 
The second is the computational difficulty due to the strong increase of CPU time and memory demand with increased propagation time which limits the 
spectral resolution and the duration of nonequilibrium calculations. We have recently presented an idea how to overcome the first and substantially weaken the second problem: invoking the generalized Kadanoff--Baym ansatz (GKBA) of Lipavsk\'y~\textit{et 
al.}~[\onlinecite{lipavsky86}]. This concept was tested on the level of second order Born selfenergies for the example
of a one-dimensional (1D) Hubbard cluster containing just two sites and two electrons because here comparisons with available 
exact diagonalization methods are easily possible, cf.~[\onlinecite{hermanns_jpcs13}, \onlinecite{hermanns12_pysscripta}, \onlinecite{bonitz_cpp13}].

Based on these encouraging first results, in this paper, we present a systematic analysis  of the HF-GKBA approach in application to Hubbard nanoclusters.
We discuss fundamental questions such as the issue of total energy conservation and we extend our previous results to larger systems. This issue has regained importance since the idea to use the GKBA in NEGF calculations has recently been taken up 
for charge dynamics in molecular junctions [\onlinecite{stefanucci_gkba}] and the dynamics of localization in 1D Hubbard chains [\onlinecite{lev_14}]. While Bar Lev et al. [\onlinecite{lev_14}] used Hartree--Fock propagators
Stefanucci et al.~[\onlinecite{stefanucci_gkba}] used the GKBA in combination with Hartree--Fock as well as damped propagators. Here, we demonstrate that only the use of Hartree--Fock propagators (HF-GKBA) retains the conserving properties of the original selfenergy. The constraints on the propagators are reformulated giving rise to a generalization and relaxation of the well-known energy conservation criteria of Baym and Kadanoff \cite{baym61,book_kadanoffbaym_qsm}.

We, furthermore, report excellent numerical behavior (long-time stability of the propagation) of the HF-GKBA and confirm the advantageous scaling with the propagation time ($\sim$\,$T^2$, compared to $T^3$ in full two-time NEGF simulations [\onlinecite{hermanns12_pysscripta}]). Comparing to exact diagonalization results, we conclude that the HF-GKBA with second Born selfenergies is very accurate for weak coupling, for a time duration substantially larger than that of time-dependent Hartree--Fock and that this time range increases with the particle number.
Finally, we use the HF-GKBA to study the short-time dynamics of Hubbard clusters and present applications to larger systems.

This paper is 
organized as follows: In Sec. \ref{s:negf}, we recall the basics of the nonequilibrium Green functions approach. The 
transition to a single-time description with the help of the GKBA is discussed in Sec.~\ref{s:gkba}, where we also discuss 
its conserving character, in dependence on the choice of the propagators. Our numerical results for finite Hubbard clusters are presented in Sec.~\ref{s:res}.

% ------------- THEORY -------------------------------
\section{Nonequilibrium Green functions}\label{s:negf}
\begin{figure}
\includegraphics[width=8.cm]{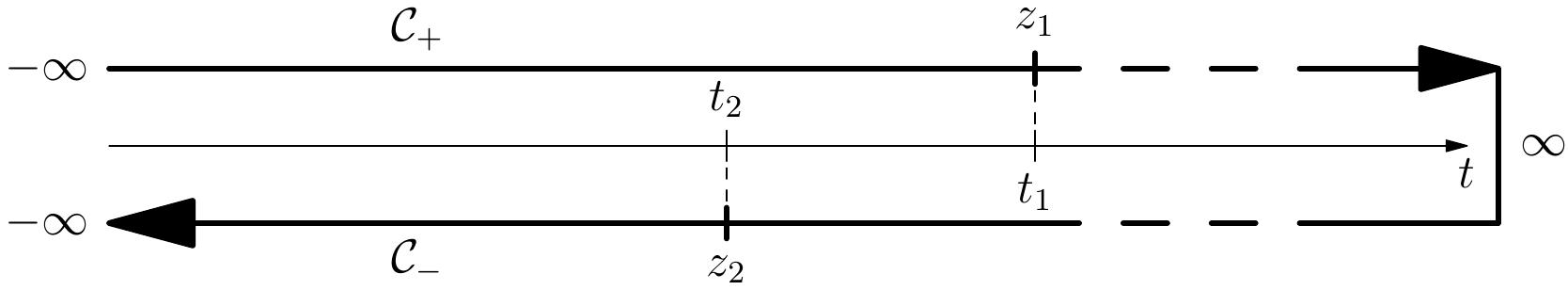}
\caption{Schwinger/Keldysh contour $\mathcal{C}$. The $\mathcal{C}_+$-branch starts from $-\infty$ running in positive 
direction up to $\infty$, the $\mathcal{C}_-$-branch starts from $\infty$ and leads in negative direction to $-\infty$. 
Note that on the contour the marked time $z_2$ is later than $z_1$, although the corresponding times on the real time 
axis satisfy $t_1>t_2$.}
\label{fig:kontur} 
\end{figure}

To describe correlation effects and excitations in quantum many-particle systems, the NEGF approach has proven very 
successful, as it allows for a systematical inclusion of correlations by diagrammatic expansions. In contrast to density 
matrix based schemes, the Green function method additionally offers direct access to the spectral information as well 
as particle removal and addition energies.    
The main quantity is the one-particle Green function, defined as (we set $\hbar~\equiv~1$)
\begin{align}\label{eq:g-def}
G(z,z') = -\mathrm{i}\avg{\mathcal{T}_\mathcal{C}\left[\Psi(z)\Psi^\dagger(z')\right]},
\end{align}
where the brackets denote thermodynamic averaging and $\mathcal{T}_\mathcal{C}$ is the time ordering operator on the 
Schwinger--Keldysh round trip contour~[\onlinecite{KeldyshContour}] $\mathcal{C}$, on which the times $z$ and $z'$ are defined, 
see Fig.~\ref{fig:kontur}.
$\Psi^{(\dagger)}$ denotes a one-particle annihilation (creation) operator in an arbitrary one-particle basis in 
second quantization. To simplify the notation we suppress the orbital index ``i'' regarding $\Psi^{(\dagger)}$ as 
vectors $\Psi^{(\dagger)}_i \rightarrow \Psi^{(\dagger)}$. Correspondingly, the Green function (\ref{eq:g-def}) is 
understood as a matrix with components $G_{ij}$.

For a quantum system of $N$ particles in nonequilibrium, the generic time-dependent Hamiltonian is given by
\begin{equation}
 \op{H}(t) =  \sum_{i=1}^N {\hat h}_i(t) + \frac{1}{2}\sum_{i\ne j}^N {\hat V}_{ij}, \quad {\hat h}_i(t) = \frac{{\hat p}_i^2}{2m}+{\hat U}_i(t).
\label{eq:h_general}
\end{equation}
In second quantization, this expression attains the form
\begin{equation}
 \op{H}(t) =  \Psi^\dagger(t) h(t) \Psi(t) + \frac{1}{2} \Psi^\dagger(t) \Psi^\dagger(t) V(t) \Psi(t)\Psi(t),
\end{equation}
where $h(t)$ denotes the one-particle Hamiltonian including an external potential,  and $V(t)$ is an arbitrary 
(possibly time-dependent) pair-interaction potential [here $h$ and $V$ are understood as matrices $h_{ij}$  and 
$V_{ijkl}$, respectively]. 

The equations of motion for $G$ are the first equation of the Martin--Schwinger hierarchy~[\onlinecite{martin_theory_1959}] and 
its adjoint,
\begin{align}
\label{eq:KBE}
\Big[i\partial_{z}-h(z)\Big]G(z,z')&=\delta_\mathcal{C}(z-z') \\
&\quad+\int_\mathcal{C} d\bar z\,W(z,\bar z)\,G^{(2)}(z\bar z;\,z' \bar z^+) \,,
\nonumber\\
\label{eq:KBE_ad}
G(z,z')\left[-i\partial^{z'}-h(z')\right]&=\delta_\mathcal{C}(z-z')\\
&\quad+ \int_\mathcal{C} d\bar z\,W(z',\bar z)\,G^{(2)}(z\bar z^-;\,z' \bar z),\nonumber
\end{align}
where $W(z,z') = V(z)\delta_C(z-z')$, and $\delta_C$ is a delta function on the contour ${\cal C}$.
The Martin--Schwinger hierarchy is equivalent to the exact many-body problem, describing the coupling of the evolution 
of the one-particle Green function to the two-particle Green function, 
\begin{equation}\label{eq:g2}
G^{(2)}(z_1z_2;\,z'_1z'_2)=-\avg{\mathcal{T}_\mathcal{C}\left[
\Psi(z_1)\Psi(z_2)\Psi^\dagger(z'_2)\Psi^\dagger(z'_1)\right]},
\end{equation}
which itself is coupled to the three--particle Green function by a similar equation (the Bethe--Salpeter equation) and so 
on. 

\begin{figure}
  \includegraphics[width=8.cm]{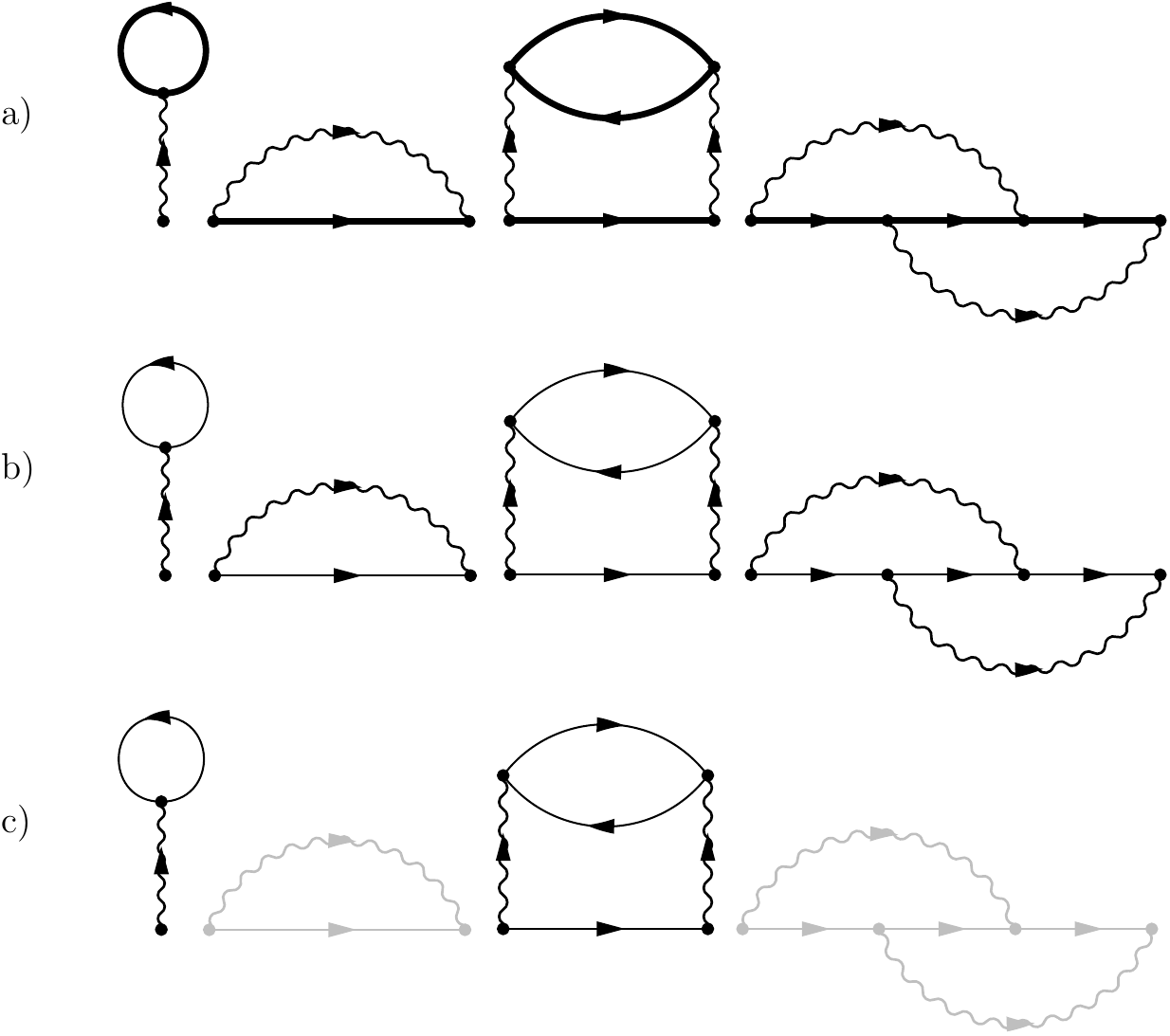}
\caption{(a) Set of nonequilibrium Feynman diagrams used in the present
work, from left: Hartree, Fock and second order Born (direct and
exchange) selfenergy. Wiggly lines denote the interaction potential,
full lines are two-time Green functions. (b) When the HF--GKBA is
applied, all full lines are replaced via Hartree--Fock Green functions
Eqs.~(\ref{eq:GKBA}, \ref{eq:propagators}), (thin lines). (c)
Non-vanishing diagrams used in the HF-GKBA calculations for the Hubbard
clusters. The exchange diagrams (gray lines) do not appear.}
\label{fig:diagrams_sigma}
 \end{figure}

To solve this hierarchy and to make it numerically tractable, a formal decoupling is performed by introducing the self-energy $\Sigma(z,z')=\Sigma[G(z,z')]$ which is a functional of the single-particle Green function. This self-energy can be found from a diagrammatic expansion in terms of Feynman diagrams, where only some classes of diagrams are chosen according to the properties of the examined system.
With this, the equations of motion (\ref{eq:KBE}) and (\ref{eq:KBE_ad}) become formally closed equations for $G$:  
\begin{align}
\left[\mathrm{i}\partial_{z}-h(z)\right]G(z,z')&= \delta_\mathcal{C}(z-z')+ \nonumber\\
& \quad\int_\mathcal{C} d\bar z\,\Sigma[G](z,\bar z)G(\bar z,z')\,, 
%\nonumber\\
\label{eq:KBESigma}\\
G(z,z')\left[-i\partial^{z'}-h(z')\right]&= \delta_\mathcal{C}(z-z') +  \nonumber\\
& \quad\int_\mathcal{C} d\bar z\,G(z,\bar z)\Sigma[G](\bar z,z')\,,
%\nonumber\\
\label{eq:KBESigma_ad}
\end{align}
which are the Keldysh--Kadanoff--Baym equations (KBE).
The KBE are---in principle---exact equations of motion of the many-body system would the selfenergy be exactly known. This is the case only for a limited number of models. In general, therefore, one has to resort to many-body approximations for the selfenergy. Baym and Kadanoff have shown how to select approximations that obey the conservation laws [\onlinecite{baym61,book_kadanoffbaym_qsm}], which we will discuss in Sec.~\ref{s:gkba}. 
The simplest approximation is the Hartree--Fock (HF) approximation where correlations are neglected entirely. It is commonly expected that this is a reasonable approximation for weak coupling. Nevertheless, we will see that, even for small coupling strength, in some nonequilibrium situations correlation effects may play a crucial role, in particular, for the long-time behavior. Among the higher order selfenergies, we mention the second(-order) Born (2B), GW or T-matrix approximations [\onlinecite{book_bonitz_qkt}]. In this paper, we will focus on the second order Born approximation shown diagramatically in Fig.~\ref{fig:diagrams_sigma}, as it allows for long time simulations. Note that, due to the particular nature of the interaction in the Hubbard model, the Fock and the 2B-exchange diagrams (gray, second and fourth) vanish. For the treatment of Hubbard nanoclusters in higher order approximations, we refer to Ref.~[\onlinecite{puigvonfriesen10}].
The remarkable property of the NEGF theory is that, via utilization of the roundtrip contour [\onlinecite{KeldyshContour}], all approximations known from ground state theory and thermodynamic equilibrium situations remain fully valid in nonequilibrium, including slow and rapid processes as well as weak and strong excitation. 

The direct numerical solution of the KBE (\ref{eq:KBESigma},\ref{eq:KBESigma_ad}) is now routine, e.g. [\onlinecite{book_bonitz_qkt}, \onlinecite{book_bonitz_semkat}, \onlinecite{balzer_lnp13}, \onlinecite{stefanucci_book_13}] and references therein. After preparing a correlated initial state, e.g. [\onlinecite{semkat_jmp00}, \onlinecite{stefanucci13}], the system is propagated in the two-time plane by computing the NEGF as a function of both time arguments. Due to the time-memory structure of the collision integral in Eqs.~(\ref{eq:KBESigma}, \ref{eq:KBESigma_ad}), the NEGF at all times and for all values of the orbital (site and spin) indices have to be stored in memory [\onlinecite{fedvr}]. Here, substantial advances could be recently achieved via a sophisticated program structure and parallelization~[\onlinecite{balzer_pra10,balzer_pra10_2}]. Nevertheless, the computational requirements for the KBE solutions exhibit an unfavorable cubic scaling with time [\onlinecite{hermanns12_pysscripta}]. Clearly, this limits the duration of propagation in nonequilibrium as well as the accuracy and resolution of the computed energy spectra that are obtained from a Fourier transform (time integral over the whole simulation). To overcome these limitations and make the long-time calculations feasible, we will apply the generalized Kadanoff--Baym ansatz (GKBA) which leads to a quadratic scaling with time and also has a number of other attractive features. This ansatz is discussed in detail in the next section.

%------------------
\section{The generalized Kadanoff--Baym ansatz}\label{s:gkba}
The one-particle Green functions appearing in Eqs.~(\ref{eq:KBE}) and (\ref{eq:KBE_ad}) depend on two times $z, z'$ both of which can be located on either the upper or lower branch of the contour ${\cal C}$, cf. Figure \ref{fig:kontur}. We note that we do not use a contour with a third (vertical imaginary) branch to produce a correlated initial state, e.g. [\onlinecite{stefanucci13}], [\onlinecite{balzer_lnp13}]. Instead, we will prepare this state by an initial real-time propagation during which the interaction is turned on adiabatically, e.g. [\onlinecite{rios11,hermanns12_pysscripta}]. Thus $G(z,z')$ on ${\cal C}$ represents a $2\times 2$ matrix, i.e. 4 functions with two conventional real time arguments $t, t'$. Two of these functions are independent. It is common to use the following definitions for the correlation ($>, <$) and retarded (R) and advanced (A) functions:
\begin{eqnarray}
   G_{ij}^<(t,t') &=& \mathrm{i} \avg{\Psi^\dagger_j(t')\Psi_i(t)},\\
   G^>_{ij}(t,t') &=& -i \avg{\Psi_i(t)\Psi^\dagger_j(t')},\\
   G^{\tn{R/A}}_{ij}(t,t')  &=& \pm \Theta[\pm(t-t')]\left[ G^>_{ij}(t,t') - G^<_{ij}(t,t')  \right].\quad 
\label{eq:gr_g<-g>}
%\hspace*{-0.12cm}.
\end{eqnarray}
To make the relations to the field operators clear, we temporarily restored the orbital indices $i,j$. In the following, these indices will be suppressed again, i.e. $G$ and $\Sigma$ have to be understood as matrices $G_{ij}$ and $\Sigma_{ij}$.
The equations of motion for the correlation functions $G^{\gtrless}$ and the propagators $G^\tn{R/A}$ follow directly from the KBE (\ref{eq:KBESigma}, \ref{eq:KBESigma_ad}) on the contour ${\cal C}$, applying the Langreth-Wilkins rules, e.g. [\onlinecite{LWR}, \onlinecite{balzer_lnp13}]. For the correlation functions, we have
\begin{eqnarray}
\label{eq:kbe_less} 
&&\Big[\mathrm{i}\partial_{t}-h(t)\Big]G^<(t,t') =  \\\nonumber
& \qquad & \int d\bar t\,\left\{ \Sigma^{\tn{R}}(t,\bar t)G^<(\bar t,t') + \Sigma^<(t,\bar t)G^{\tn{A}}(\bar t,t')\right\}\,, 
\\
&&G^<(t,t')\Big[-\mathrm{i}\partial^{t'}-h(t)\Big] =  \nonumber\\\nonumber
& \qquad & \int d\bar t\,\left\{ G^{\tn{R}}(t,\bar t)\Sigma^<(\bar t,t') + G^<(t,\bar t)\Sigma^{\tn{A}}(\bar t,t')\right\}\,, 
\\
\label{eq:kbe_gtr} 
&&\Big[\mathrm{i}\partial_{t}-h(t)\Big]G^>(t,t') =  \\\nonumber
& \qquad & \int d\bar t\,\left\{ \Sigma^\tn{R}(t,\bar t)G^>(\bar t,t') + \Sigma^>(t,\bar t)G^\tn{A}(\bar t,t')\right\}
\\
&&G^>(t,t')\Big[-\mathrm{i}\partial^{t'}-h(t)\Big] =  \nonumber
\\\nonumber
& \qquad & \int d\bar t\,\left\{ G^{\tn{R}}(t,\bar t)\Sigma^>(\bar t,t') + G^>(t,\bar t)\Sigma^{\tn{A}}(\bar t,t')\right\}\,,  
\end{eqnarray}
where the components ``R/A'' and ``$\gtrless$'' of $\Sigma$ are defined analogously to those of G.
The propagators satisfy 
\begin{eqnarray}
\label{eq:kbe_ra} 
&&\Big[\mathrm{i}\partial_{t}-h(t)\Big]G^\tn{R/A}(t,t') = \delta(t-t')+ \\\nonumber
& \qquad & \int d\bar t\,\Sigma^\tn{R/A}(t,\bar t)G^\tn{R/A}(\bar t,t')\,, \\\nonumber
&&G^\tn{R/A}(t,t')\Big[-\mathrm{i}\partial^{t'}-h(t)\Big] = \delta(t-t')+ \\\nonumber
& \qquad & \int d\bar t\,G^\tn{R/A}(t,\bar t)\Sigma^\tn{R/A}(\bar t,t')\,. 
\end{eqnarray}
Lipavsk\'y et al. [\onlinecite{lipavsky86}] have shown that the KBE for $G^<$ are equivalent to an integral equation
($t>t'>t_0$):
\begin{align}
\label{eq:G_R_less}
G^<(t,t') &= -G^R(t,t')\rho(t')\\
&+ \int_{t'}^{t}d\bar{\bar t}\,\int_{t_0}^{t'}d \bar t\, G^R(t,\bar{\bar t})\Sigma^<(\bar {\bar t}, \bar t)G^A(\bar t, t')\nonumber\\ &+\int_{t'}^{t}d\bar{\bar t}\,\int_{t_0}^{t'}d \bar t\,G^R(t,\bar {\bar t})\Sigma^R(\bar {\bar t}, \bar t)G^<(\bar t, t')\,,\nonumber 
\end{align}
whereas for $t_0<t<t'$:
\begin{align}
\label{eq:G_A_less}
G^<(t,t') &= \rho(t)G^A(t,t')\\
&- \int_{t}^{t'}d\bar{\bar t}\,\int_{t_0}^{t}d \bar t\, G^R(t,\bar{\bar t})\Sigma^<(\bar {\bar t}, \bar t)G^A(\bar t, t')\nonumber\\ &-\int_{t}^{t'}d\bar{\bar t}\,\int_{t_0}^{t}d \bar t\,G^<(t,\bar {\bar t})\Sigma^A(\bar {\bar t}, \bar t)G^A(\bar t, t')\,. \nonumber
\end{align}
Note, that by exchanging ($<\Lra>$) and replacing the density matrix $\rho=:f^<$ by $f^>=1-f^<$ in equations (\ref{eq:G_R_less}) and (\ref{eq:G_A_less}), the analogous expression for $G^>$ is easily obtained. Details of the derivation can be found in Ref.~[\onlinecite{hermanns12_pysscripta}]. 

\subsection{GKBA}
While expressions (\ref{eq:G_R_less}) and (\ref{eq:G_A_less}) are still exact (within the chosen approximation for $\Sigma$) they contain the unknown two-time function $G^<$ also under the integral on the r.h.s. Therefore, one can attempt to solve the integral equation up to second order, approximating $G^<$ under the integral just by the first term: 
\begin{align}
\label{eq:GKBA}
G^\gtrless_{{\tn{GKBA}}}(t,t')=-G^R(t,t')f^\gtrless(t')+f^\gtrless(t)G^A(t,t').
\end{align}
This is the generalized Kadanoff--Baym ansatz (GKBA) of Lipavsk\'y, \ifmmode \check{S}\else \v{S}\fi{}pi\ifmmode \check{c}\else \v{c}\fi{}ka and Velick\'y which is exact on the time diagonal, $t=t'$.
The importance of this equation lies in the fact that it provides a means for the reconstruction---though approximately---of the off-diagonal Green functions (i.e. for arguments $t\ne t'$) from single-time quantities such as the density matrix $f^<$. Note that the argument of $f^\gtrless$ is not the mean of the two times appearing on the left but always the earlier of the two times. This means the GKBA retains the retardation structure (memory) of the collision integrals and thus obeys causality, which turns out to be crucial for the conservation properties, see Sec.~\ref{ss:econs}.

Note that Eq.~(\ref{eq:GKBA}) indicates that the GKBA is only formally closed in terms of $\rho(t)$, since it still involves two-time quantities---the retarded (advanced) propagators $G^R\,(G^A)$. These functions obey equations of motion of similar complexity as $G^{\gtrless}(t,t')$, cf. Eq. (\ref{eq:kbe_ra}). To make further progress, one can use approximate propagators that are not computed selfconsistently with $G^{\gtrless}(t,t')$.
\subsection{Hartree--Fock GKBA}\label{ss:hf-gkba}
In this paper, we approximate the propagators by Hartree--Fock propagators, $G^\tn{R/A} \to G^\tn{R/A}_\tn{HF}$ that are obtained from the solution of 
the KBE (\ref{eq:kbe_ra}) with the replacement $\Sigma^\tn{R/A} \to \Sigma^\tn{R/A}_\tn{HF}$, with the solution ($\mathcal{T}$ is the causal time-ordering operator):
\begin{equation}
\label{eq:propagators}
G^\tn{R/A}_\tn{HF}(t,t')=\mp\mathrm{i}\Theta[\pm(t-t')]\mathcal{T}\left[\exp\left(-\mathrm{i}\int_{t'}^td \bar t\,H(\bar t)\right)\right],
\end{equation}
where $H$ denotes the Hartree--Fock (mean--field) Hamiltonian, 
\begin{eqnarray}
H(t) &=& h(t) + \Sigma_{\rm HF}(t),\\
 &=:& h_0 + {\bar \Sigma}_{\rm HF}(t)
\label{eq:h_hf}
\end{eqnarray}
 that contains the time-dependent external field [via $h(t)$] and interaction effects via the Hartree--Fock (mean field) selfenergy $\Sigma_{\rm HF}(t)$ that selfconsistently involves the time--dependent density matrix. For later purposes, we also introduced the definition ${\bar \Sigma}_{\rm HF}$ for the sum of Hartree--Fock selfenergy and external field, with  $h_0$ being just the stationary single-particle energy (kinetic plus potential energy).

Approximation (\ref{eq:propagators}), together with the GKBA, Eq.~(\ref{eq:GKBA}), will be called Hartree--Fock GKBA (HF--GKBA). One motivation of this choice is that, in the special case that the system is treated within the HF approximation (neglecting the correlation contribution, $\Sigma_\tn{cor}=0$), the HF--GKBA is exact, i.e. Eqs. (\ref{eq:GKBA}, \ref{eq:propagators}) provide the exact solution for $G^{\gtrless}$ which is readily confirmed by direct solution of the KBE (\ref{eq:kbe_less}, \ref{eq:kbe_gtr}) in Hartree--Fock.
Note that HF--GKBA can be used for an arbitrary correlation selfenergy $\Sigma_\tn{cor}$. The example of the T-matrix selfenergy will be briefly discussed in Sec.~\ref{s:dis}.

Let us now have a closer look at the HF--GKBA for the case that correlations are taken into account and discuss its consequences. The Dyson equation for the full Green function on the contour ${\cal C}$ can be written as
\begin{eqnarray}\label{eq:dyson_gkba}
G &=& G_\tn{id} +  G_\tn{id} \left( {\bar \Sigma}_{\rm HF} + \Sigma_\tn{GKBA} + \Delta \Sigma \right) G,
\end{eqnarray}
where $G_\tn{id}$ denotes the ideal Green function of the uncorrelated and field-free system.
As discussed in Ref.~[\onlinecite{bonitz_cpp99_2}], this equation can be decomposed into several coupled integral equations that allow to construct the full Green function in steps. Here, we choose to split-off the correlation selfenergy and introduce the HF-Green function according to
\begin{eqnarray}\label{eq:dyson_gkba1}
G_{\rm HF} &=& G_\tn{id} +  G_\tn{id} {\bar \Sigma}_{\rm HF} G_{\rm HF}, 
\\\label{eq:dyson_gkba2}
G_{\rm GKBA} &=& G_{\rm HF} +  G_{\rm HF} \Sigma_{\rm GKBA} G_{\rm GKBA}, 
\\
\label{eq:dyson_gkba3}
G &=& G_{\rm GKBA} +  G_{\rm GKBA} \Delta\Sigma \;G.
\end{eqnarray}
We defined $\Sigma_{\rm GKBA}$ as the correlation selfenergy in which all two-time functions are reconstructed according to the HF--GKBA, Eqs. (\ref{eq:GKBA}, \ref{eq:propagators}), thus, $\Sigma_{\rm GKBA} \equiv \Sigma_\tn{cor}[f^\gtrless, G^\tn{R/A}_{\rm HF}]$. The deviation of the full selfenergy from this approximation has been denoted $\Delta \Sigma \equiv \Sigma_\tn{cor} - \Sigma_{\rm GKBA}$ and contains terms with one, two and three full propagators and $G^\tn{R/A}$, respectively. Thus, the Green function computed within the HF--GKBA is given by Eq.~(\ref{eq:dyson_gkba2}), containing renormalizations by the Hartree--Fock selfenergy and the second Born selfenergy with propagators on the Hartree--Fock level. In contrast, the full Green function $G$ that is computed by a full two-time calculation (i.e. without the GKBA) has undergone another renormalization given by Eq.~(\ref{eq:dyson_gkba3}). In addition to $G_{\rm GKBA}$, the function $G$ contains contributions from $\Delta \Sigma$ to all orders, so the difference of the two is given by the infinite series
\begin{eqnarray}
\label{eq:dyson_gkba4}
G-G_{\rm GKBA} &=& G_{\rm GKBA} \Delta\Sigma \:G_{\rm GKBA} \\\nonumber
&+& G_{\rm GKBA} \Delta\Sigma \:G_{\rm GKBA} \Delta\Sigma \:G_{\rm GKBA} + \dots,
\end{eqnarray}
where each subsequent term adds contributions with up to three full (renormalized by correlations) propagators $G^\tn{R/A}$.
% The two different approximations---$G$ versus $G_{\rm GKBA}$---are illustrated in Fig.~\ref{fig:diagrams_sigma} by the full lines [part (a)] and thin 
% lines [part (b)], respectively.
 
The properties of the GKBA have been studied for macroscopic spatially homogeneous systems [\onlinecite{bonitz_jpcm96}]. There, it was found that the GKBA retains the conservation laws of the original two-time approximation for the selfenergy [\onlinecite{book_bonitz_qkt, bonitz_pla96}]. This will be considered in more detail in Sec.~\ref{ss:econs}. As to the accuracy of the GKBA, it was found that this ansatz is a very good approximation to the full two-time solution if the exact propagators $G^\tn{R/A}(t,t')$ are being used, indicating that the additional integral contributions in Eqs.~(\ref{eq:G_R_less}) and (\ref{eq:G_A_less}) are often of minor importance. In the case of approximate propagators, good results were obtained with ideal as well as Hartree--Fock propagators [\onlinecite{kwong98}]. In contrast, the use of damped propagators that include imaginary selfenergy contributions, violates total energy conservation and leads to an overall worse performance [\onlinecite{bonitz_epjb99}]. The use of the HF--GKBA for finite Hubbard clusters [\onlinecite{hermanns12_pysscripta,bonitz_cpp13}] confirms these results. Details will be presented in Sec.~\ref{s:res}.
\subsection{Conservation of total energy}\label{ss:econs}
%\subsubsection{NEGF criterion of a conserving approximation}
The issue of conserving approximations is of central importance for the treatment of correlated many-body systems. This is, in particular, relevant for the many-body dynamics far from equilibrium. It is an attractive feature of Green functions theory that conserving approximations are straightforwardly selected.
Baym and Kadanoff [\onlinecite{baym61,book_kadanoffbaym_qsm}] have formulated a simple criterion for a NEGF approximation to be conserving that consists of two conditions: 
\begin{description}
 \item[A] the single-particle Green function obeys simultaneously the KBE and its adjoint, Eqs.~(\ref{eq:KBE}) and (\ref{eq:KBE_ad}), and
 \item[B] the two-particle Green function is symmetric with respect to both particles, i.e. \\
$G^{(2)}(1,2;1',2') = G^{(2)}(2,1;2',1')$.
\end{description}
These conditions easily allow one to select conserving approximations for the two-particle Green function and the selfenergy.

We now show that, when a conserving approximation for $\Sigma$ is being used, the subsequent application of the HF--GKBA does not change the conservation properties. Application of the GKBA amounts to solving the KBE only along the time diagonal, $z=z'$. The corresponding equation of motion for $G(z,z)$ is 
obtained by computing the difference of the KBE  and its adjoint, Eqs. (\ref{eq:KBE}, \ref{eq:KBE_ad}), cf. Ref.~[\onlinecite{book_kadanoffbaym_qsm}],
\begin{eqnarray}\nonumber
&& \left\{ \i\left(\partial_{z_1}+\partial_{z'_1} \right) - \left[ H(z_1)-H(z'_1) \right]\right\} G(z_1,z_1')\big|_{z_1=z_1'} = 
%\qquad\qquad
\\
&&\qquad \qquad \pm \i \int_{\cal C} dz_2 \left[ V(z_1-z_2) - V(z'_1-z_2)\right] 
\nonumber\\
&& \qquad\qquad \qquad \qquad G_{\rm cor}^{(2)}(z_1,z^-_2,z'_1,z^+_2)\big|_{z_1=z_1'},
\label{eq:kbe_diagonal}
\end{eqnarray}
where, in the end, $z_1'=z_1$ is set, and we introduced the correlation part of the two-particle Green function, $G_{\rm cor}^{(2)} \equiv G^{(2)} - G_{\rm HF}^{(2)}$ [recall that $H$ contains the Hartree--Fock selfenergy, cf. Eq.~(\ref{eq:h_hf})]. By construction, the solution $G(z,z)$ fulfills condition A. 

Consider now condition B. To this end we use the solution for the correlation part of $G^{(2)}$ that follows from the Bethe-Salpeter equation. In what follows, it will be sufficient to consider the screened ladder approximation (SCA), Ref.~[\onlinecite{bornath99,bonitz_jpcs13}]
\begin{eqnarray}
 G_{\rm cor}^{(2)}(z_1,z_2,z'_1,z_2') &=& \i \int_{\cal C} d{\bar z}_1 d{\bar z}_2 G(z_1{\bar z}_1)G(z_2{\bar z}_2) \times
\nonumber\\\label{eq:g2_bse}
&& \quad V({\bar z}_1,{\bar z}_2)G^{(2)}({\bar z}_1,{\bar z}_2,z'_1,z_2').\quad
\end{eqnarray}
This equation can be solved by iteration, starting by replacing 
$G^{(2)}(z_1,z_2,z'_1,z_2') \longrightarrow G^{(2)}_{\rm HF}(z_1,z_2,z'_1,z_2')=G(z_1z'_1)G(z_2z'_2)\pm G(z_1z'_2)G(z_2z'_1)$ under the integral. This first iteration corresponds to the dynamically screened second Born approximation (GW) which, obviously, is symmetric in the labels of particles 1 and 2, in agreement with condition B. If we now apply the HF--GKBA to each Green function under the integral, this symmetry is fully retained. Thus, we have shown that the HF--GKBA for the GW approximation is conserving. The same applies to the static limit when $V(z_1, z_2) \to V(z_1)\delta_c(z_1 - z_2)$, i.e. the static second Born approximation is conserving as well when the HF--GKBA is applied. The same proof applies to the T-matrix approximation and to the SCA. To show this, we only need to proceed further with the iterative procedure for the solution of Eq.~(\ref{eq:g2_bse}), either with the static or dynamic potential. It is easy to realize that each term of the iteration series has the needed symmetry $1 \longleftrightarrow 2$, resulting in the fulfillment of condition B.

Thus, we conclude that the application of the GKBA to an arbitrary conserving approximation of NEGF theory does not change the exchange symmetry, condition B. This symmetry is also retained when, in addition, the HF-approximation for the propagators (\ref{eq:propagators}) is made, and our numerical results for the HF-GKBA fully confirm total energy conservation. There is, however, a serious problem with the previous argument. Let us consider damped propagators, i.e. replace 
$G^{\rm R/A}_{\rm HF}(t,t') \rightarrow G^{\rm R/A}_{\rm HF}(t,t') \exp[-\gamma(t-t')]$. This approximation is known to violate energy conservation [\onlinecite{bonitz_jpcm96,bonitz_epjb99}], although it also clearly obeys the symmetry $1 \longleftrightarrow 2$ and, thus, fulfills conditions A and B.
%Evidently, the conserving conditions A and B of Baym and Kadanoff are necessary but not sufficient. 
In order to understand the origin of the violation 
of energy conservation for the exponentially damped propagators we, therefore, now first consider a different approach that is based on density operator theory.
We will then return to the conditions A and B in Sec.~\ref{ss:general_consreving} and resolve this contradiction.

\subsection{Density operator theory and the GKBA}\label{ss:do-gkba}
With the application of the HF--GKBA to the KBE, the problem becomes a closed non-Markovian equation for the time-diagonal element of the Green function, i.e. for the one-particle density matrix. Such an equation can, of course, be derived independently from a one-time theory of reduced density matrices. This was first shown in Ref. [\onlinecite{book_bonitz_qkt}], see also Refs. [\onlinecite{bonitz_jpcm96,hermanns_jpcs13}]. This equivalence is important to identify the HF--GKBA with standard approximations from density operators. At the same time, results from density operator theory, including conservation laws and long-time behavior, can be used to analyze the properties of the single-time solutions of the KBE. Finally, we note recent interest in density operator methods in the context of the relaxation dynamics of finite Hubbard clusters [\onlinecite{akbari_12}].

We, therefore, briefly recall the concept of the reduced nonequilibrium density operators 
\begin{eqnarray}\label{eq:fs}
F_{1\ldots s} &=& \frac{N!}{(N-s)!}\tn{Tr}_{s+1\ldots N}\rho_{N}\,,\quad
\\\nonumber
 \tn{Tr}_{1\ldots s}F_{1\ldots s} &=& \frac{N!}{(N-s)!}\,,
\end{eqnarray}
where $\rho_N$ is the density operator of the full system which is normalized to unity and $F_{1\dots s}$ is the 
associated $s$-particle operator. The equations of motion for the $F_{1\dots s}$ (BBGKY-hierarchy) follow from 
the von Neumann equation for $\rho_N$ by taking the partial trace,
\begin{eqnarray}
\nonumber
\mathrm{i}\hbar \partial_t\, F_1-\Comm{h_1\,,F_1} &=& \tn{Tr}_2\Comm{V_{12}\,,F_{12}}\,, \\
\mathrm{i}\hbar \partial_t\, F_{12}-\Comm{H_{12}\,,F_{12}}&=& \tn{Tr}_3\Comm{V_{13}+V_{23}\,,F_{123}}\,,
\nonumber\\
\dots & \dots & \dots \nonumber\\
\label{eq:bbgky}
\end{eqnarray}
where the two-particle hamiltonian is $H_{12}=h_1+h_2+V_{12}$, and
the system (\ref{eq:bbgky}) has to be complemented by initial conditions for $F_1, F_{12}$ etc. The equations are coupled and form a hierarchy that eventually stops when the r.h.s. involves $F_{1\dots N}=\rho_N$, in analogy to the two-time Martin--Schwinger hierarchy of NEGF, see Sec.~\ref{s:negf}. As in the case of NEGF, the 
hierarchy is usually decoupled at a low level by replacing the exact $F_{1\dots s}$ by an approximation $F^\tn{app}_{1\dots s}[F_1,...F_{1\dots s-1}]$ that is a functional of the lower order operators. Key approximations of NEGF theory, including the second Born approximation [\onlinecite{bonitz_jpcm96}], ladder approximation  [\onlinecite{kremp_ap97}] or GW approximation [\onlinecite{book_bonitz_qkt}] are readily identified by proper choices for the three-particle density operator, see also Ref. [\onlinecite{hermanns_jpcs13}]. Since the density operator approach does not involve two-time Green functions, full agreement with NEGF theory requires, in addition, the time-diagonal limit as provided by the GKBA, and in fact the GKBA is directly recovered in the theory of reduced density operators, e.g. [\onlinecite{book_bonitz_qkt}]. For the present purpose of analyzing energy conservation, it is sufficient to note that the HF-GKBA is directly recovered from the system (\ref{eq:bbgky}). On the other hand, the GKBA with correlated propagators containing correlation selfenergy contributions $\Sigma^{\rm cor}$ [such as exponential damping] leads to a modification of the second hierarchy equation by the replacement 
\begin{align}
{\hat H}_{12} &\rightarrow {\hat H}^{\rm eff}_{12} = {\hat H}_{12} + {\hat \Sigma}_{12}^{\rm cor},
\label{eq:h12_sigma}
\\
\Comm{H_{12}\,,F_{12}} &\rightarrow H^{\rm eff}_{12} F_{12} - F_{12} H^{\rm eff \dag}_{12}
\nonumber
\end{align}
with all other terms left unchanged compared to the HF-GKBA.

It is this renormalization of the two-particle hamiltonian that destroys the conservation of total energy in the GKBA with correlated propagators. To show this, 
we recall the derivation of conserving approximations in density operator theory [\onlinecite{dufty_boercker,book_bonitz_qkt}], for related issues of density operator theory, we refer to Refs. [\onlinecite{dufty_boercker89,dufty_97}].
We begin with expressing the mean kinetic, potential and interaction energy via the one-particle and two-particle density operators,
\begin{eqnarray}
 \langle {\hat T} \rangle &=& \tn{Tr}_{1} \frac{{\hat p}^2_1}{2m} F_1, \qquad \langle {\hat U} \rangle = \tn{Tr}_{1} {\hat U}_1 F_1,
\label{eq:mean_tu}
\\
\langle {\hat V} \rangle &=& \frac{1}{2}\tn{Tr}_{12} {\hat V}_{12} F_{12}.
\label{eq:mean_v}
\end{eqnarray}
We now compute the time derivative of kinetic energy using the first hierarchy equation,
\begin{align}
& \i\hbar \frac{d}{dt} \langle {\hat T} \rangle =
\tn{Tr}_{1}  \frac{{\hat p}^2_1}{2m}\Comm{h_{1},F_1} + \frac{1}{2}\tn{Tr}_{12} \frac{{\hat p}^2_1+{\hat p}^2_2}{2m} \Comm{{\hat V}_{12},F_{12}}
\nonumber
\\
&= -\tn{Tr}_{1} {\hat U}_1 \Comm{h_{1},F_1} + \frac{1}{2}\tn{Tr}_{12} \left(H_{12}-{\hat U}_1-{\hat U}_2 \right) \Comm{{\hat V}_{12},F_{12}}
\label{eq:dtdt}
\end{align}
The time derivative of the potential energy follows similarly,
\begin{align}\label{eq:dudt}
 \i\hbar \frac{d}{dt} \langle {\hat U} \rangle &=
\i\hbar \tn{Tr}_{1}  \frac{\partial {\hat U}_1}{\partial t} F_1 + \\
& \tn{Tr}_{1}  {\hat U}_1\Comm{h_{1},F_1} + \frac{1}{2}\tn{Tr}_{12} \left( {\hat U}_1+{\hat U}_2 \right) \Comm{{\hat V}_{12},F_{12}}.
\nonumber
\end{align}
Finally, the time derivative of the interaction energy is transformed using the second hierarchy equation with the replacement (\ref{eq:h12_sigma})
\begin{align}\label{eq:dvdt}
 \i\hbar \frac{d}{dt} \langle {\hat V} \rangle &=
 \frac{1}{2}\tn{Tr}_{12}  \left( H^{\rm eff}_{12} F_{12} - F_{12} H^{\rm eff \dag}_{12}\right)
\\
&+ \frac{1}{2}\tn{Tr}_{123} V_{12} \Comm{{\hat V}_{13}+{\hat V}_{23},F_{123}}.
\nonumber
\end{align}

Collecting the results (\ref{eq:dtdt}, \ref{eq:dudt}, \ref{eq:dvdt}), we obtain, for the time derivative of the total energy $H=T+U+V$ minus the 
power introduced into the system by the external potential,
\begin{align}\label{eq:dhdt}
 \frac{d}{dt}\langle H\rangle -\tn{Tr}_{1}  \frac{\partial {\hat U}_1}{\partial t} F_1 &= \frac{1}{2\i\hbar}\tn{Tr}_{123} V_{12} \Comm{{\hat V}_{13}+{\hat V}_{23},F_{123}}
\\\nonumber
& \quad +  \frac{1}{2\i\hbar}\tn{Tr}_{12}  \left( \Sigma^{\rm cor}_{12} F_{12} - F_{12} \Sigma^{\rm cor \dag}_{12}\right).
\end{align}
%where $\Sigma_{12}^{\rm cor}=\Sigma^{\rm cor}_1+\Sigma^{\rm cor}_2$.
For a conserving approximation, the right hand side has to be equal to zero. The first term vanishes if the 
 three-particle density operator is symmetric with respect to the particle indices, $F_{123}(t)=F_{132}(t)=F_{321}(t)$, at all times. This is an obvious and trivial condition [it is similar to condition B of NEGF theory for the two-particle Green function, cf. Sec.~\ref{ss:econs}. Here, in the case of single-time operators, it appears on the three-particle level], and is fulfilled also for the exact solution. 

To verify this symmetry for the second Born approximation, we first rewrite $F_{123}$ in terms of correlation operators (cluster expansion) and give the corresponding expression for the pair-correlation operator, details can be found in Refs.~[\onlinecite{book_bonitz_qkt}]~and~[\onlinecite{hermanns_jpcs13}],
\begin{align}
 F_{123} &= \Lambda_{123}^{\pm} \left\{ F_1F_2F_3 + F_1c_{23} + F_2c_{13} + F_3c_{12} \right\}, 
 \nonumber
\\
 \Lambda_{123}^{\pm} &= \frac{1}{3!}\left\{1 \pm P_{12} \pm P_{13} \pm P_{23} +  P_{12}P_{13} + P_{13}P_{23}  \right\}
 \nonumber\\
 \mathrm{i}\hbar \partial_t\, c_{12} &-\Comm{{\bar H}^0_{12}\,,c_{12}} = {\hat V}_{12}F_1F_2  - F_2F_2 {\hat V}^{\dagger}_{12}\,,
%\nonumber
\end{align}
where ${\hat V}_{12} = (1 \pm F_1 \pm F_2)V_{12}$, $\;{\bar H}^0_{12} = {\bar H}^0_{1}+ {\bar H}^0_{2}$ is the two-particle Hartree--Fock hamiltonian and $\Lambda_{123}^{\pm}$ is the three-particle (anti-)symmetrization operator.
Obviously, this set of equations assures symmetry of $F_{123}$ in the particle indices. This approximation is equivalent to the HF-GKBA in second Born approximation, i.e. no renormalization of the two-particle hamiltonian (\ref{eq:h12_sigma}) is performed.

As noted above, for the HF-GKBA, the second term on the r.h.s. in Eq.~(\ref{eq:dhdt}) is absent and energy conservation is confirmed for any conserving approximation of two-time NEGF theory. In contrast, for the GKBA with propagators that contain correlation selfenergies, $\Sigma^{\rm cor}$, even for a conserving approximation (with a properly symmetric $F_{123}$) energy conservation is destroyed. In fact, just the anti-hermitean part of $\Sigma^{\rm cor}$, which governs the damping behavior of the propagators, contributes to the second term on the r.h.s. This is particularly evident in the Born approximation where $\Sigma_{12}^{\rm cor}=\Sigma^{\rm cor}_1+\Sigma^{\rm cor}_2$, and each single-particle propagator $G_1^{\rm R/A}$ is renormalized by a single-particle correlation selfenergy contribution $\Sigma^{\rm cor}_1$.

\subsection{Energy conservation of the GKBA revisited. Relaxing the conservation criteria of Baym and Kadanoff}\label{ss:general_consreving}  
After having confirmed energy conservation for the HF-GKBA and violation thereof for damped propagators within an independent density operator approach, it remains to establish how this result can be obtained within NEGF theory. In particular, the question arises whether conditions A and B of Baym and Kadanoff are indeed sufficient and necessary for energy conservation.

We first find out why the symmetry of $G_{12}$ alone (condition B) does not imply energy conservation of the GKBA for arbitrary choice of the propagators as it appeared in Sec.~\ref{ss:econs}. Let us go back to condition A. The strict meaning of the  statement (implied, by Baym and Kadanoff) that the Green function $G(z,z')$ obeys simultaneously the KBE and its adjoint, Eqs.~(\ref{eq:KBE} ,\ref{eq:KBE_ad}) is that {\em each of the real-time Keldysh components} $G^>, G^<, G^R, G^A$  simultaneously fulfills these equations. Since only two of these functions are independent it is sufficient to consider $G^<$ and $G^R$ which obey the pairs of equations (\ref{eq:kbe_less}) and (\ref{eq:kbe_ra}), respectively. In these equations, the two-particle Green function is eliminated in favor of the selfenergy Keldysh matrix with the same components $\Sigma^{\gtrless}$ and $\Sigma^{R/A}$ and the same link between them
\begin{align}
   \Sigma^{\tn{R/A}}_{ij}(t,t')  &= \pm \Theta[\pm(t-t')]\left[ \Sigma^>_{ij}(t,t') - \Sigma^<_{ij}(t,t')  \right].
\label{eq:sigma_ra_gl}
\end{align}

In our argument in Sec.~\ref{ss:econs} we used the condition that $G^<$ fulfills its pair of KBE. But what about $G^R$? In standard two-time NEGF theory, of course, also $G^R$ fulfills its pair. But this is no longer the case when we apply the GKBA. In this approximation, the standard relation between $G^{\gtrless}$ and $G^{R/A}$ is altered and replaced by Eq.~(\ref{eq:GKBA}). This means, $G^{R/A}$ do not follow from the result for $G^>$ and $G^<$ but obey an independent equation. In other words, the equation of motion for $G^{R/A}$, if written again in the standard form (\ref{eq:kbe_ra}), will contain a selfenergy $\Sigma^{\tn{R/A}}$ that is independent of $\Sigma^{\gtrless}$ and not given by Eq.~(\ref{eq:sigma_ra_gl}). In fact, the GKBA just uses this independence in favor of a simpler choice for $\Sigma^{\tn{R/A}}$ to simplify the calculations. For example, the HF-GKBA uses just $\Sigma^{\tn{R/A}} = \Sigma^{\rm HF}$, whereas the approximation with the exponentially damped propagators is associated with $\Sigma^{\tn{R/A}} = \Sigma^{\rm HF} - i \gamma$ where $\gamma$ is time-independent. Finally, we can establish a connection with the corresponding approximation for the two-particle Green function from the standard mapping $\Sigma(1,1') G(1,1') \longleftrightarrow \tn{Tr}_{2} W(1,2)G^{(2)}(12;1'2)$. While the Hartree--Fock selfenergy corresponds to $G^{(2)}_{\rm HF}$ which is obviously symmetric in the particle indices, the term $\gamma G_1$ is associated with a non-symmetric function $G^{(2)}$. This explains the preservation of energy conservation for the HF-GKBA and its violation for the GKBA with exponentially damped propagators where both statements are independent of the original choice of $\Sigma$ provided it was conserving.

We may now use this result to revise the conditions A and B such that they apply to the GKBA. In fact we can relax the conditions of the theorem of Baym and Kadanoff (BKT) such that they cover not only the GKBA but a broader class of conserving approximations than envisaged originally in Refs.~[\onlinecite{baym61, book_kadanoffbaym_qsm}].\\[6ex]
{\bf Theorem:} The time-dynamics of a system described by the single-particle Green functions $G^<$ and $G^R$ connected via a functional relation $G^<=G^<_{\rm ansatz}[G^R]$, with 
\begin{description}
 \item[A1] $G^R$ obeying simultaneously the real-time KBE involving the selfenergy $\Sigma_{I}$, Eq.~(\ref{eq:kbe_ra1}),  and its adjoint, 
  \begin{eqnarray}
  \label{eq:kbe_ra1} 
  &&\Big[\mathrm{i}\partial_{t}-h(t)\Big]G^\tn{R/A}(t,t') = \delta(t-t')+ \\\nonumber
  & \qquad & \int d\bar t\,\Sigma_I^\tn{R/A}(t,\bar t)G^\tn{R/A}(\bar t,t')\,, 
  \end{eqnarray}
 \item[A2] $G^<$ obeying simultaneously the real-time KBE involving the selfenergy $\Sigma_{II}$, Eq.~(\ref{eq:kbe_less1}), and its adjoint,
  \begin{eqnarray}
  \label{eq:kbe_less1} 
  &&\Big[\mathrm{i}\partial_{t}-h(t)\Big]G^<(t,t') =  \\\nonumber
  & \qquad & \int d\bar t\,\left\{ \Sigma_{II}^{\tn{R}}(t,\bar t)G^<(\bar t,t') + \Sigma_{II}^<(t,\bar t)G^{\tn{A}}(\bar t,t')\right\}\,, 
 \end{eqnarray}
 \item[B] and the two-particle Green functions associated with $\Sigma_I$ and $\Sigma_{II}$ being symmetric with respect to both particles, i.e. \\
$G_I^{(2)}(1,2;1',2') = G_I^{(2)}(2,1;2',1')$ and \\
$G_{II}^{(2)}(1,2;1',2') = G_{II}^{(2)}(2,1;2',1')$,
\end{description}
is conserving.

{\bf Proof:} Consider first Eq.~(\ref{eq:kbe_ra1}). This equation is decoupled from $G^<$. Thus, the symmetry condition B for $G_I$ guarantees, according to the BKT, that
 $\Sigma^{R/A}_I$ and, hence the dynamics of $G^R$, are conserving. Consider now Eq.~(\ref{eq:kbe_less1}). It contains $\Sigma_{II}$ which, according to condition B, is conserving. Eq.~(\ref{eq:kbe_less1}), in addition, contains $G^R$ and $G^A$. Since the dynamics of $G^{R/A}$ is conserving and the functional relation 
$G^<_{\rm ansatz}[G^R]$ is a single-particle relation having the same form for $G^<(1,1')$ and $G^<(2,2')$, we conclude that the coupling to $G^{R/A}$ does not destroy the conservation properties of $\Sigma_{II}$. Thus, the problem is reduced to the original BKT, and the dynamics of $G^<$ is conserving as well what proves the theorem.

Thus, instead of one selfenergy there are now two selfenergies that have to be conserving simultaneously. Obviously, this version of the theorem reduces to the BKT if the connection between $G^<$ and $G^R$ is given by the standard NEGF relation, Eq.~(\ref{eq:gr_g<-g>}) and, consequently, $\Sigma_I\equiv \Sigma_{II}$. If the relation $G^<_{\rm ansatz}[G^R]$ is given by Eq.~(\ref{eq:GKBA}) we recover the GKBA where the choice of the retarded propagators is given by $\Sigma_I$. Finally, there exists a whole set of new conserving approximations ($\Sigma_I, \Sigma_{II}$) where the functional relation (\ref{eq:GKBA}) is replaced by a different one, 
$G^<_{\rm ansatz}[G^R]$.

\section{Results for finite Hubbard clusters}\label{s:res}
Let us now apply the NEGF formalism with the HF--GKBA to the dynamics of standard finite lattice systems described by the Hubbard model. The purpose of this section is three-fold:
\begin{enumerate}
 \item  to test conceptual questions of the GKBA and the numerical performance. It has been reported before that two-time NEGF simulations for small   
   Hubbard clusters exhibit unphysical damping behavior [\onlinecite{friesen_10}]. Furthermore, approximations based on density matrices have shown numerical instabilities in the long-time regime [\onlinecite{akbari_12}]. We will show in Sec.~\ref{ss:gkba-test} that these problems do not occur with the HF--GKBA,
  \item to extend the simulations to larger systems where no exact diagonalization data are available, cf. Sec.~\ref{ss:gkba-test}, and 
  \item to investigate the dynamical relaxation behavior of small Hubbard clusters at different coupling strengths, cf. Sec.~\ref{ss:cor-dyn}.
\end{enumerate}

\subsection{Hubbard model and second Born approximation}\label{ss:hubbard}
We consider a finite Hubbard model with $N_b$ sites at half-filling (particle number $N=N_b$) with hopping amplitude $J$ and on-site interaction $U$.
The initial Hamiltonian, for times $t<0$, reads
\begin{align}
 \op{H}_{0} &=-J\sum_{<s,s'>}\sum_{\sigma=\uparrow,\downarrow}\op{c}^\dagger_{s,\sigma}\op{c}_{s',\sigma}+{\tilde U}\sum_{s}\op{n}_{s,\uparrow}\,\op{n}_{s,\downarrow}\;
+ 
\nonumber\\
& \qquad + \sum_{ij,\alpha\beta}{\tilde f}_{ij,\alpha\beta}(t)\,\op{c}_{i\alpha}^\dagger\op{c}_{j\beta},
\label{eq:h0}
\end{align}
where $s$ and $s'$ label the discrete sites, and $<\!s,s'\!>$ indicates nearest-neighbor sites. Further, $\op{n}_{s,\sigma}=\op{c}^\dagger_{s,\sigma}\op{c}_{s,\sigma}$ denotes the density operator. The last term in Eq.~(\ref{eq:h0}) incorporates an external time-dependent excitation that drives the system out of equilibrium. Several cases for the choice of the function ${\tilde f}$ will be specified below. In the following, we will use dimensionless parameters where energy (time) is measured in units of $J$ (the inverse hopping amplitude $J^{-1}$) and the coupling strength and field amplitude will be given by $U={\tilde U}/J$ and 
$f={\tilde f}/J$, respectively.

In our simulations, we start from an non-interacting initial state and adiabatically turn on the interactions to reach a fully correlated state. After this, the external excitation is turned on. Details on the procedure can be found in Refs. [\onlinecite{rios11,hermanns_jpcs13,bonitz_cpp13}].
We have verified in advance that the switching is slow enough to avoid any artifacts in the dynamics. In practice, we use a Fermi-like switching function 
\begin{equation}
f(t) = 1-\left[1 + \exp\left(\frac{t-t_f/2}{\tau}\right)\right]^{-1} 
\end{equation}
with a switching duration $t_f=50$ and a switching constant $\tau=3$, which yields sufficiently converged results.
\subsection{Testing the GKBA for small clusters}\label{ss:gkba-test}
\begin{figure}
    \includegraphics{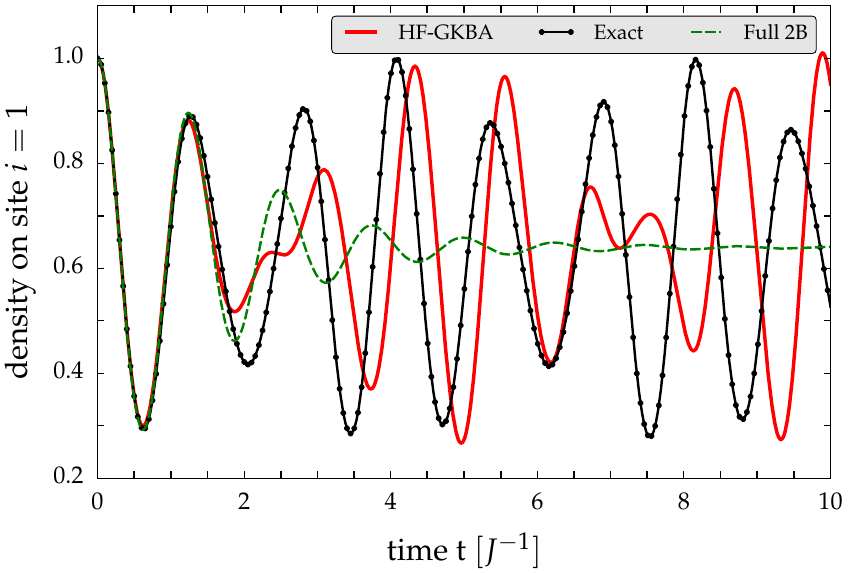}\\
    \includegraphics{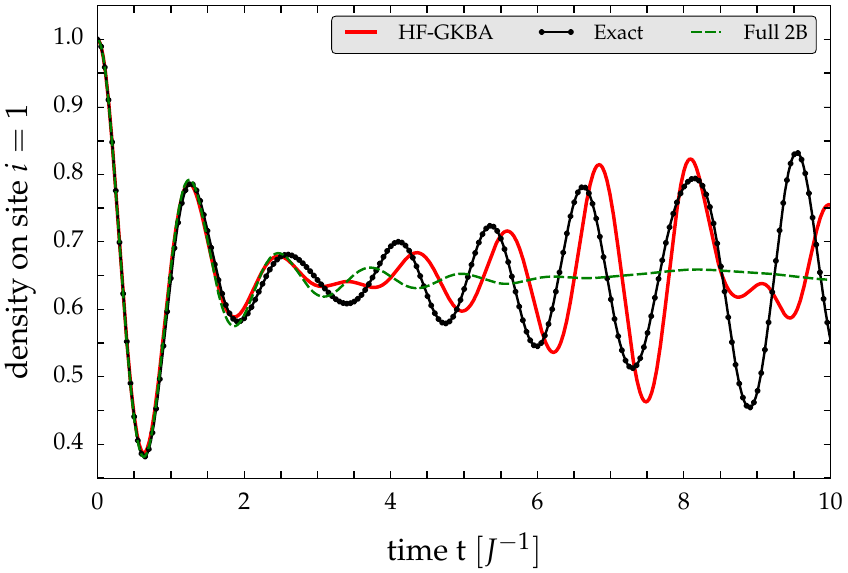}
\caption{(Color online) {\bf Top.} Time evolution of the density on site $1$ for a 2-site Hubbard model at half filling and $U=1.0$ within HF--GKBA (full red line), compared to the exact result (full black line with dots) and a two-time second Born calculation (green dashed line). The HF--GKBA does not exhibit artificial damping, in contrast to the two-time result. 
{\bf Bottom.} The same for a cluster of $N_\tn{b} = 8$ sites.
}
\label{fig:verdozzi-vergleich}
\end{figure}
We start by considering small one-dimensional Hubbard clusters where exact diagonalization results are available. This allows for a rigorous test of the HF--GKBA results in its full time dependence and for a broad variety of excitation conditions.

Let us first investigate the problem of artificial damping of the dynamics of Hubbard clusters that was observed in full two-time KBE simulations when the system was driven far out of equilibrium [\onlinecite{puigvonfriesen09}]. 
In that reference, at time $t \ge 0$, a two-site Hubbard cluster at half filling is strongly perturbed by a rapid change of the energy of site ``1'', which leads to the following choice for the last term in the Hamiltonian (\ref{eq:h0}), cf. Ref. [\onlinecite{balzer_jpcs13}],
$f_{ij,\alpha\beta}(t)=w_0\delta_{i,1}\delta_{j,1}\delta_{\alpha,\beta}\Theta(t)\,$, where $w_0=5.0$.
The results are shown in Fig.~\ref{fig:verdozzi-vergleich}. One clearly sees the rapid damping of the density on the perturbed site, in the two-time simulation in second Born approximation, while the exact result exhibits a non-decaying dynamics. 
The HF--GKBA, interestingly, does not exhibit the damping of the two-time result. Since the many-body approximations are in both cases identical, the difference is solely due to the HF--GKBA and the broken selfconsistency, as was discussed in Sec.~\ref{ss:hf-gkba}, cf. Fig.~\ref{fig:diagrams_sigma}. We note that we observe quantitative deviations from the exact result which, however, become substantially smaller when the particle number increases, cf. the figures below. The removal of the artificial damping is an important generic feature of the HF--GKBA and was confirmed in all our simulations. The same behavior is observed when the T-matrix selfenergy is used, see Sec.~\ref{s:dis}.
This gives us confidence for applying this approximation to Hubbard clusters in nonequilibrium, in particular to larger systems where no exact diagonalization data are available.
 \begin{figure}
  \includegraphics{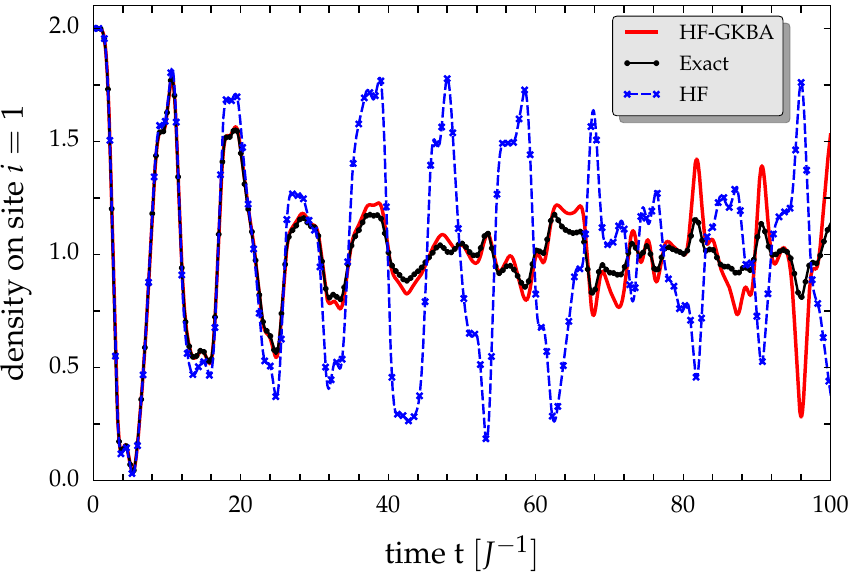}\\
  \includegraphics{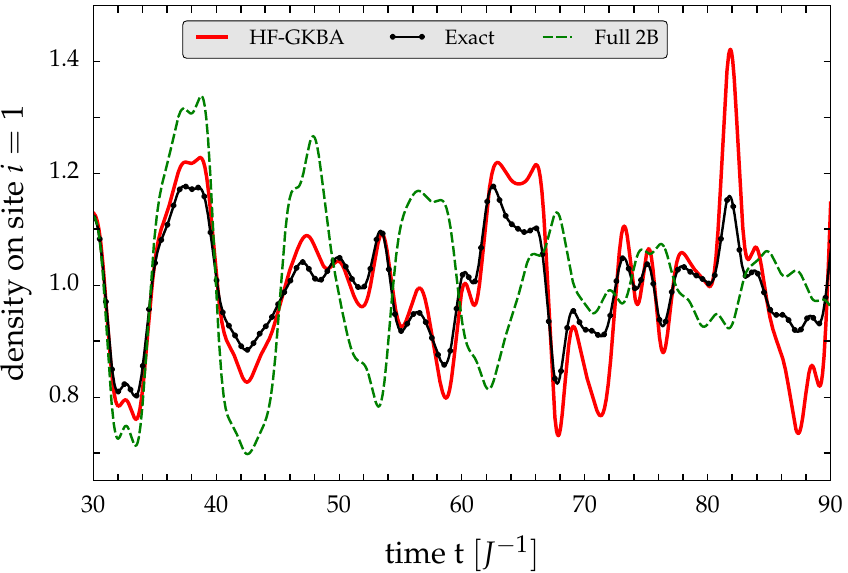}
\caption{(Color online) {\bf Top.} Density evolution on the left-most site for an 8-site Hubbard model at half filling and coupling strength $U=0.1$. Initially all particles were on the leftmost four sites. The HF--GKBA result (full red line) is compared to the exact one (full black line with dots) and to Hartree--Fock (dashed blue line with crosses). {\bf Bottom.} Same as above, but for a reduced time interval. Instead of TDHF, a two-time second Born calculation (green dashed line) is included.}
\label{fig:akbari_88_u01}
 \end{figure}

Let us now turn to larger clusters and a more quantitative comparison with exact data. 
Figure \ref{fig:akbari_88_u01} shows the dynamics of 8 electrons in an 8-site Hubbard chain at weak coupling, $U=0.1$, starting from a strong nonequilibrium situation where all particles are confined to the four left-most sites by applying a strong confinement potential. At time $t=0$, this potential is removed instantaneously and the dynamics are followed (this scenario was studied in Ref.~[\onlinecite{akbari_12}]). One observes very strong oscillations of the site occupations as particles move to the right, towards the originally empty four sites. After about four periods, these oscillations become anharmonic and continue with a reduced amplitude. A Hartree--Fock calculation fails already after about two periods (compare the blue dashed curve to the exact result), indicating that the dynamics are strongly influenced by correlation effects. In contrast, our HF--GKBA approach yields very good agreement with the exact data for about 7 periods ($t\sim 60$) after which deviations are increasing, but still most of the features are reproduced qualitatively. In particular, the dominant frequencies and peak positions are still captured. Comparing the HF-GKBA to full two-time 2B results (green dashed line), we even see a better agreement with the exact solution throughout the whole simulation. 

Next, we consider the energy spectrum for this system, again at weak coupling, $U=0.1$. This is done by applying a weak very short external field pulse to site $1$, $f_{ij,\alpha \beta}(t) = w_0 \delta_{i,1}\delta_{ij}\delta_{\alpha \beta}\delta(t), \, w_0=0.01$, that excites all possible transitions. The HF--GKBA allows us to propagate the system for a long time, until $t=1000$, and to compute the time-dependence of all observables. The energy spectrum is obtained via Fourier transform of the occupation of the perturbed site, $n_1(t)$, and the results resolve about seven orders of magnitude. There are two main energy regions. For frequencies below $\omega=4$, all three approximations, including the Hartree--Fock simulation, show overall very good agreement. Only a few smaller features -- the small side peaks of the peaks around $1.3$ and $2.2$ and the little peaks around $\omega \approx 2.7, 3.2, 3.6$ are missing in the HF-calculation, indicating that these are correlation effects (most likely double exciations [\onlinecite{balzer12_epl}]). For frequencies $\omega \ge 4$, the picture changes completely. Evidently, HF misses all peaks. In contrast, the HF--GKBA performs impressively well, taking into account the very low height of the peaks in that range.
% 2-spaltig
\begin{widetext}
 \begin{figure*}\label{fig:akbari_88_u01_spec}
  \includegraphics{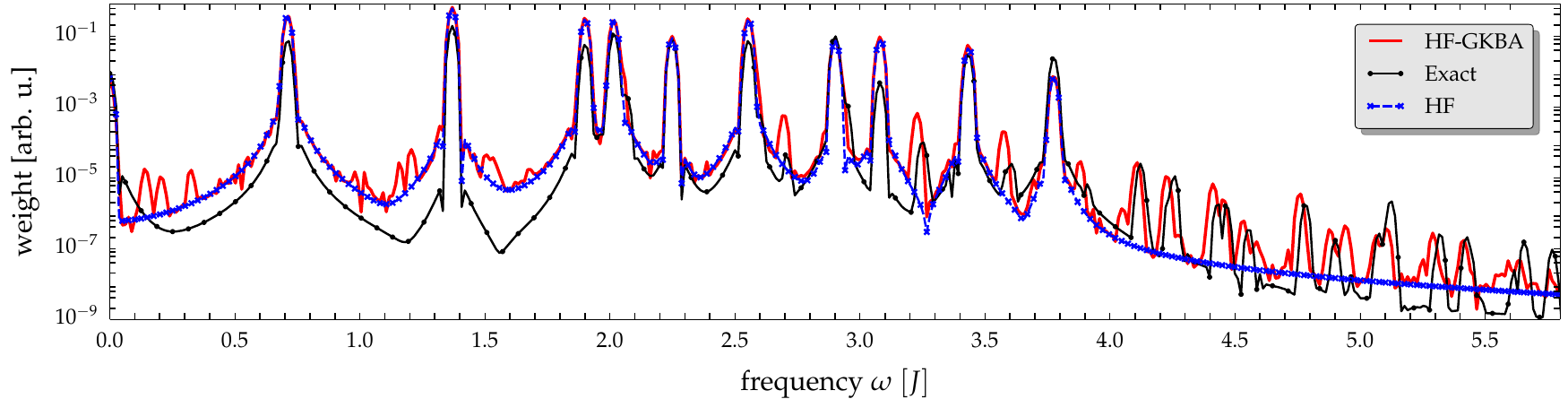}
\caption{(Color online) Energy spectrum for an 8-site Hubbard model at half filling and coupling strength $U=0.1$. The HF--GKBA result (full red line) is compared to the exact one (full black line with dots) and to Hartree--Fock (dashed blue line with crosses).}
 \end{figure*}
\end{widetext}

\begin{figure}
  \includegraphics{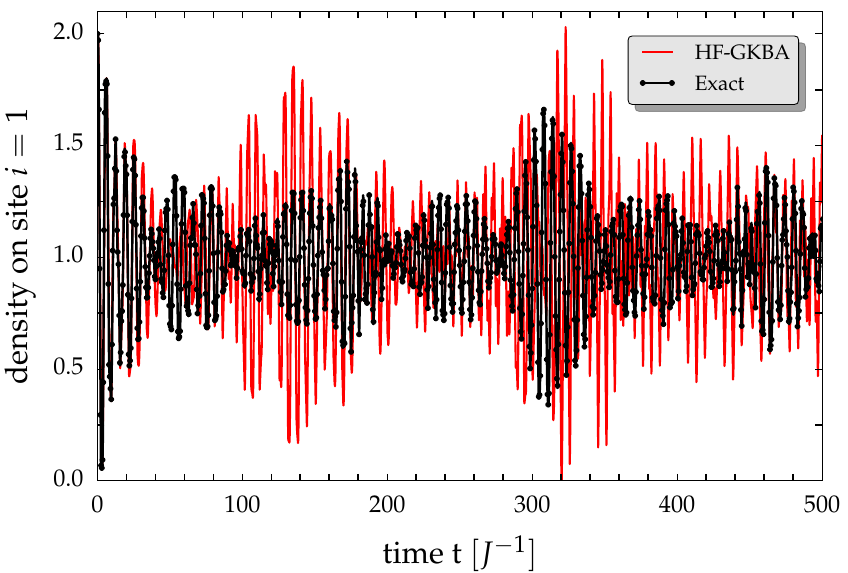}
\caption{(Color online) Long-time dynamics of a 4-site Hubbard model at half filling and $U=0.1$, following strong excitation, within HF--GKBA (red), compared to the exact result (black with dots). Initially all electrons occupy the two leftmost sites. The figures shows the time evolution of the occupation of the leftmost site.}
\label{fig:akbari_44_u01_long}
\end{figure}

After having tested the long-time behavior  of the HF--GKBA in the linear response regime, let us now consider the long-time behavior in the case of a strong non-perturbative excitation. The results for a four-site chain at half filling and $U=0.1$ are shown in Fig.~\ref{fig:akbari_44_u01_long}.
The first observation is that the simulations run stable for the entire duration of $t\le 500$. This is confirmed by additional tests that are several times longer (not shown) indicating that  previously reported stability problems [\onlinecite{akbari_12}] do not occur within the HF--GKBA. Further, the comparison with the exact result shows that the main frequencies are well reproduced, and even the phase of the oscillations is correct up to $t\sim 180$.  Yet even at longer times the overall behavior is well captured, despite the dephasing, and deviations decrease again strongly, cf., e.g., the behavior around $t=450$. At the same time, quantitative deviations are observed (amplitude of the oscillations), starting around $t\sim 90$.

\begin{figure}
    \includegraphics{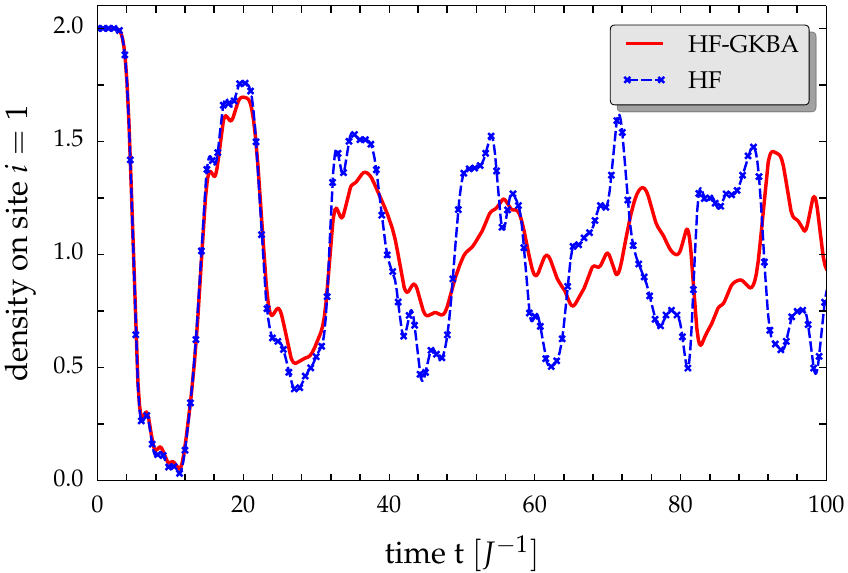}
\caption{(Color online) Time evolution of the density on the leftmost site for an 16-site Hubbard model at half filling and $U=0.1$ within HF--GKBA (full red line), compared to the HF-result (blue dashed line with crosses). Initially all particles were at the $8$ leftmost sites.}
\label{fig:akbari_n16_N16_u01_density0_big} 
\end{figure}

These observations encourage us to extend our simulations also to larger systems where no exact results are available. In all cases we confirm that the HF--GKBA is completely stable. As an example, in Fig.~\ref{fig:akbari_n16_N16_u01_density0_big}, we show the dynamics of a 16 site system at half filling and $U=0.1$ with the same type of strong nonequilibrium initial conditions (all 16 particles occupy the 8 leftmost sites).
The behavior is similar to the dynamics of the previously studied analogous systems of four and eight particles, cf. Figs.~\ref{fig:akbari_44_u01_long} and \ref{fig:akbari_88_u01}, respectively. The site occupations undergo a rapid and violent evolution which is strongly influenced by correlation effects. Hartree--Fock simulations reproduce only the first $1.5$ periods of the main oscillation ($t\lesssim 25$). Based on our HF--GKBA results, we can deduce a number of trends when the particle number is increased: first, the main oscillation period increases proportional with $N$ and, second, the oscillations become increasingly nonlinear. Since the energy spectrum becomes much more complex when the system size increases it is presently not possible to relate this oscillation to a characteristic energy transition. To shed more light onto the physical processes involved in the dynamics and the dependence on the coupling strength and particle number, we will consider an increased set of quantities below, in Sec.~\ref{ss:cor-dyn}.

Summarizing this first part of numerical results, we may conclude that the HF--GKBA is very well suited to study the dynamics of finite Hubbard clusters in the weak coupling regime, thereby (at least partially) overcoming problems of previous approaches. 
No unphysical damping and instabilities are observed. Since the present results are derived from selfenergies in second order Born approximation we have restricted the analysis to weak coupling, $U=0.1$. For moderately larger values of the coupling parameter, still acceptable results can be obtained, as will be shown below for $U=0.25$.

%--------------------------------
\subsection{Short-time dynamics: correlation build up and relaxation of site occupations}\label{ss:cor-dyn}
\begin{figure}
    \includegraphics{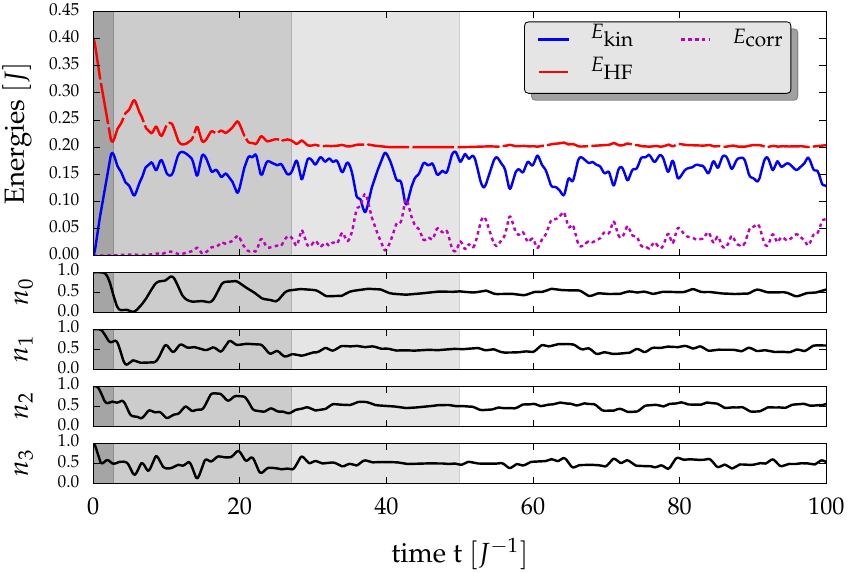}\\
    \includegraphics{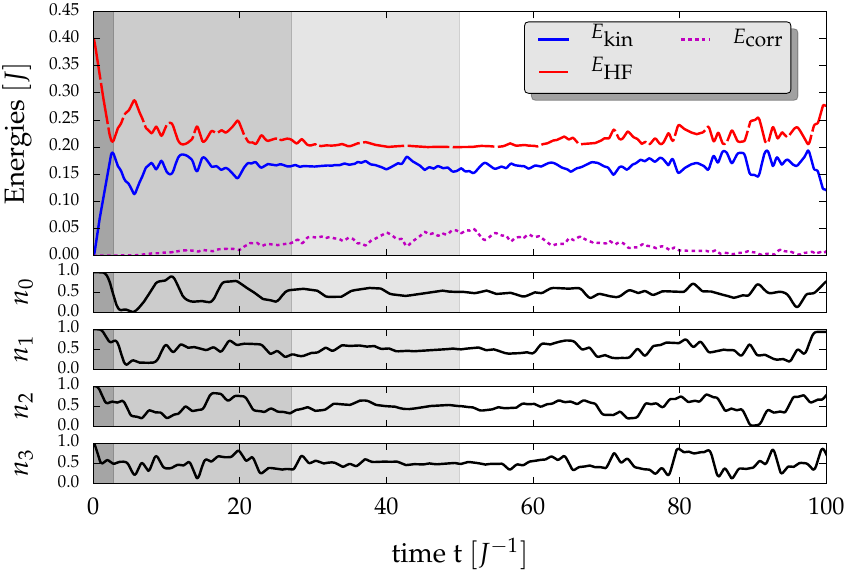}
\caption{Time evolution of the 8-site Hubbard model at half filling and $U=0.1$. Initially, all particles were at the $4$ leftmost sites. {\bf Top figure:} exact diagonalization. {\bf Bottom figure:} HF--GKBA result. Upper panels of both figures show the dynamics of kinetic, Hartree--Fock and correlation energy, bottom panels of both figures show the occupation of the four leftmost sites.
}
\label{fig:cordyn_8_01_ex} 
\end{figure}
In the following, we consider the dynamics in the same strong nonequilibrium situation that was discussed above, where all $N$ particles are initially placed on the leftmost $N/2$ sites (half filling). We now look at the behavior of additional observables. In particular, we follow the time dependence of the occupations $n_i(t)$ of all initially occupied sites (the occupations of the $N/2$ rightmost sites follow by symmetry, e.g. $n_N(t)=1-n_{1}(t)$, $n_{N-1}(t)=1-n_{2}(t)$ and so on). 

The typical dynamics can be seen in the bottom panel of Fig.~\ref{fig:cordyn_8_01_ex} displaying exact diagonalization results for the $N=8$ site chain at weak coupling, $U=0.1$. Due to Pauli blocking, initially, only electrons from site $4$ can move to the right whereas electrons from site $3$ ($2$) can only follow when site $4$ ($3$) is being depopulated. This time delay in the depopulation is clearly visible. Interestingly, the rightmost site ($4$) is only depopulated half and sites $3$ and $2$ even less. Depopulation of the leftmost site ($1$) sets in last but proceeds to the lowest value of all sites (close to zero). 
This is easy to understand: while the population of sites 2-4 is increased again by newly incoming particles from the left, no such incoming flux exists at the boundary of the chain (site 1). After a very short time, $t\approx 4$, the rightmost site ($8$) is almost fully occupied, i.e., the electron wave has reached the right border after which it is being reflected. Subsequently, a strongly nonlinear oscillatory dynamics of the site occupations occurs that is damped until the occupations reach the stationary values of a homogeneous system ($n_i=0.5, i= 1\dots 8$), around $t=50$. This, however, is not a true stationary system, as our finite system has a reversible dynamics and, consequently, we clearly observe, at later times, a strong departure from the homogeneous configuration.

It is interesting to consider, besides the occupations, also the different contributions to the total energy [note that total energy is conserved to very high accuracy] of the system which are shown in the top panel 
of Fig.~\ref{fig:cordyn_8_01_ex}. In the initial state, the system has only mean field (Hartree--Fock) energy. Both, kinetic and correlation energy are exactly zero. When the occupied sites get depopulated, we observe a rapid increase of kinetic energy which is almost completely compensated by a loss of HF energy. Kinetic energy reaches its maximum (the particle current is largest) around $t=t_{\rm I}\sim 3$ after which it decreases again and continues to oscillate. It is interesting to note that this (first) maximum of kinetic energy is reached when the population of the leftmost is decreased to $0.5$. This is observed in all simulations, independent of the coupling strength. Thus, this is a propagation effect resulting from the nonequilibrium inhomogeneous initial condition and constitutes the shortest time scale we observe in our simulations (``phase I'').

The second time scale that is apparent from the energy relaxation (``phase II'') is characterized by a formation of correlation energy and a decay and saturation of Hartree--Fock energy which is reached around $t=t_{\rm II}\sim 25$. After this time, no qualitative changes of the energy dynamics are observed, aside from nonlinear oscillations of kinetic and correlation energy that occur with almost exactly opposite time derivatives. More characteristic for this phase III is the relaxation of the site occupations which terminates around $t=t_{\rm III}\approx 50$, followed by a longer phase IV of occupation revivals, as mentioned above. In the following, we will analyze whether these four phases are visible also for other values of the coupling.

\begin{figure}
    \includegraphics{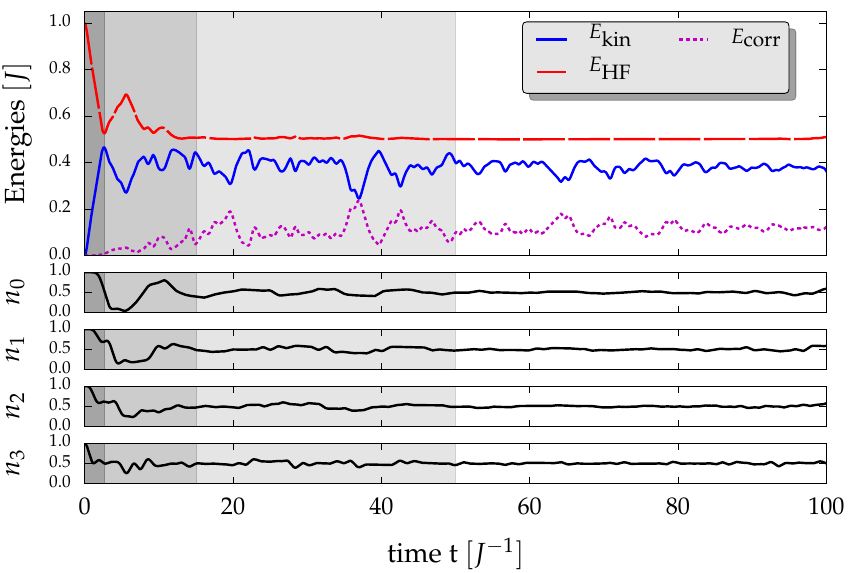}\\
    \includegraphics{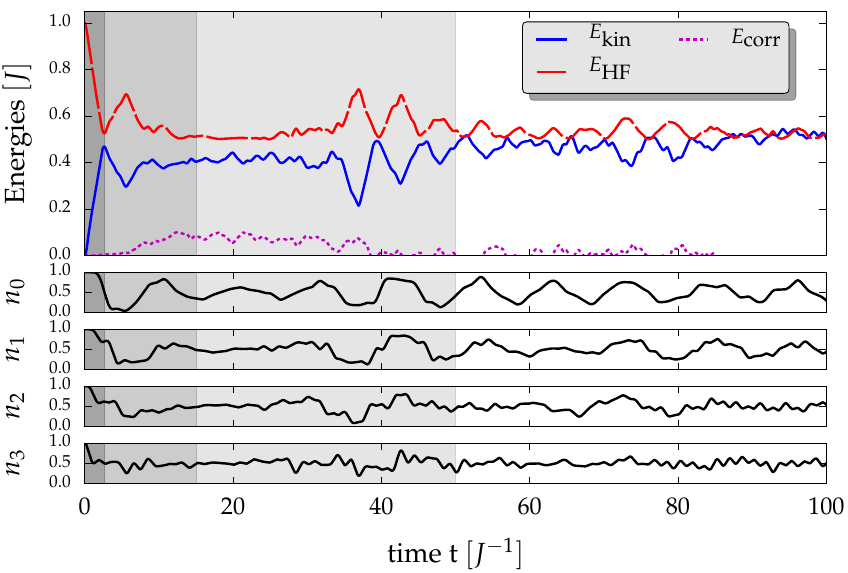}
\caption{Same as Fig.~\ref{fig:cordyn_8_01_ex} but for $U=0.25$. {\bf Top figure:} Exact diagonalization calculation. {\bf Bottom figure:} HF--GKBA}
\label{fig:cordyn_8_025_ex} 
\end{figure}

Consider now Fig.~\ref{fig:cordyn_8_025_ex}, where the dynamics for the same conditions are shown, but for the larger coupling strength $U=0.25$. Again, we observe the rapid depopulation of the sites (phase I) and associated relaxation of kinetic and Hartree--Fock energy, with the same characteristic time $t_{\rm I}\approx 3$. The main difference, compared to the case $U=0.1$, is the more rapid saturation of HF energy and build up of correlation energy (phase II) terminating around $t_{\rm II} \approx 15$. The relaxation of the occupations (phase III) lasts again until $t=t_{\rm III}\approx 50$. 
% \begin{figure}
%    \includegraphics[width=8.5cm]{./figures/akbari_n16_N16_density0_long}
%\caption{Same as Fig.~\ref{fig:cordyn_8_025_ex} but  for $N=8$ and using the HF--GKBA.}
%\label{fig:cordyn_8_025_gkba} 
%\end{figure}

\begin{figure}
    \includegraphics{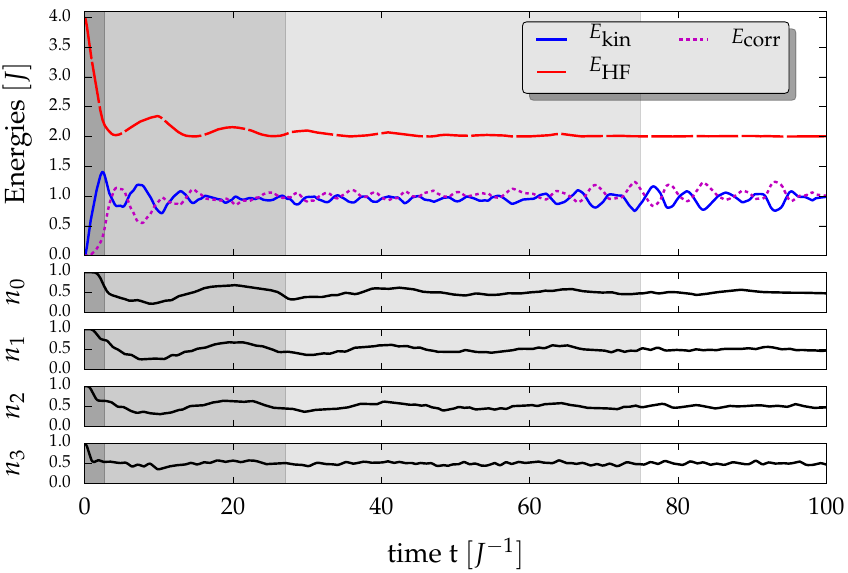}\\
\caption{Same as Fig.~\ref{fig:cordyn_8_01_ex} but for $U=1$. Exact diagonalization calculation.}
\label{fig:cordyn_8_1_ex} 
\end{figure}

Finally, we consider the case $U=1.0$, cf. Fig.~\ref{fig:cordyn_8_1_ex}. As before, phase I has a duration of 
$t_{\rm I}\approx 3$. The relaxation of the occupations now takes slightly longer, until $t=t_{\rm III}\approx 75$. The most striking difference to the previous cases, however, is in the dynamics of the correlation energy. Here, buildup of correlations is essentially over around $t=5$ whereas the Hartree--Fock energy saturates only around $t_{\rm II} \approx 15$. 

We now turn to larger particle numbers. Here, no exact diagonalization results are available and we have to resort to the HF--GKBA approximation. Based on the analysis of Sec.~\ref{ss:gkba-test}, we expect that this approximation is reliable in the case of weak coupling. This can be verified directly for the presently computed quantities by comparing the HF--GKBA dynamics for an 8-particle Hubbard chain with the exact diagonalization results. To this end, the HF--GKBA results for $U=0.1$ and $U=0.25$ are also shown in Figs.~\ref{fig:cordyn_8_01_ex} and \ref{fig:cordyn_8_025_ex}, cf. the lower figure parts. For $U=0.1$, the exact behavior of the occupations and energy contributions during phases I and II is very accurately reproduced. At the later stages deviations occur. Starting around $t=20$ the kinetic and correlation energy evolve much more smoothly than in the exact calculation. At the same time, the evolution of the site occupations is correct also during phase III, until $t\sim 60$, after which the GKBA occupations become significantly more violent than in the exact case. At $U=0.25$, the GKBA shows the correct behavior only until $t\sim 25$, i.e., with increasing coupling, deviations between exact diagonalization and HF--GKBA grow more rapidly.
\begin{figure}
    \includegraphics{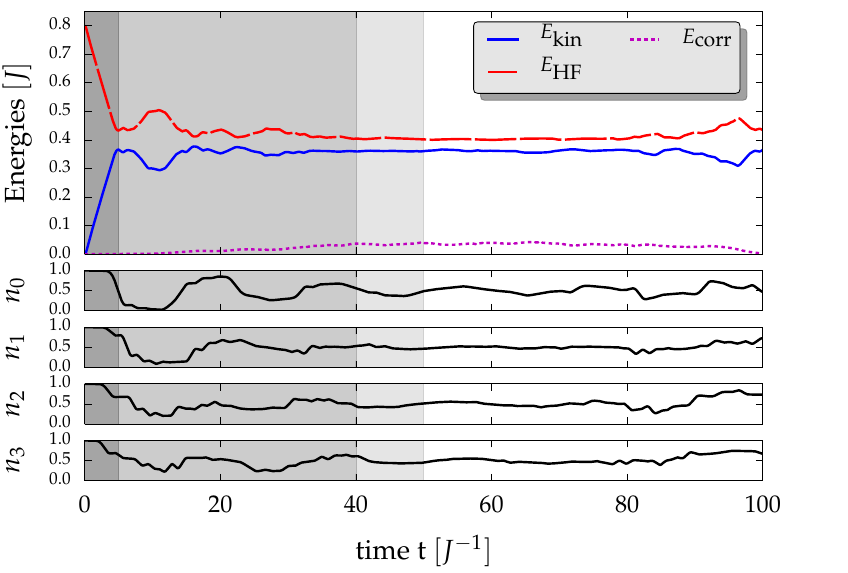}\\
\caption{Time evolution of the 16-site Hubbard model at half filling and $U=0.1$. Initially all particles were at the $8$ leftmost sites. {\bf Top panel}: dynamics of total energy (conserved), as well as of kinetic, Hartree--Fock and correlation energy. {\bf Bottom panels}: occupation of the four leftmost sites. 
 HF--GKBA result.}
\label{fig:cordyn_16_01_gkba} 
\end{figure}

\begin{figure}
    \includegraphics{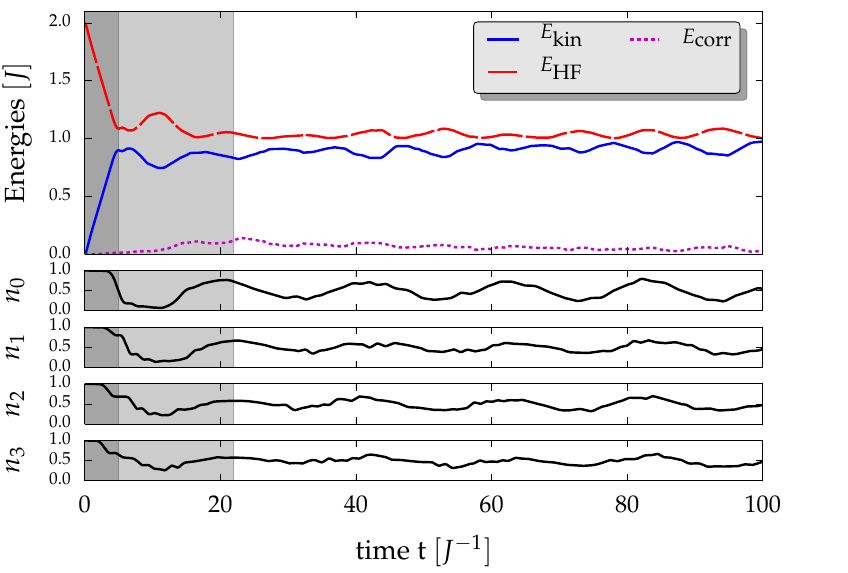}\\
\caption{Time evolution of the 16-site Hubbard model at half filling and $U=0.25$. Initially all particles were at the $8$ leftmost sites. {\bf Top panel}: dynamics of total energy (conserved), as well as of kinetic, Hartree--Fock and correlation energy. {\bf Bottom panels}: four leftmost sites. 
 HF--GKBA result.}
\label{fig:cordyn_16_025_gkba} 
\end{figure}

Based on these observations, we can now study the case of $N=16$ particles, where we limit ourselves to $U=0.1$ and $U=0.25$; the results are shown in Figs.~\ref{fig:cordyn_16_01_gkba} and~\ref{fig:cordyn_16_025_gkba}, respectively. The general observation is that the dynamics for $N=16$ are much less violent than for $N=8$, and the different stages are longer.
In particular, the decay of $n_1$ to $0.5$ and the associated increase of kinetic energy take substantially longer, until $t_{\rm I}\approx 5$. The reason is that now $8$ sites have to be depopulated (essentially sequentially), which takes longer than for four sites. Also, the build up of correlation energy and the saturation of the Hartree--Fock energy (phase II) takes longer, approximately until $t_{\rm II}\approx 40$. In contrast, the first relaxation of the populations to the homogeneous value $0.5$ takes until $t_{\rm III}\approx 50$, as in the case of $8$ particles.

For $N=16$ and $U=0.25$, we again observe a slower increase of kinetic energy (phase I, $t_{\rm I}\approx 5$) in agreement with the case $U=0.1$. Again, 
correlation energy builds up slower than for eight particles (phase II), but here the time scale appears to be shorter than for $U=0.1$ ($t_{\rm II}\approx 25$), although this result is less reliable.
%-----------------------------
\section{Summary and Discussion}\label{s:dis}
In this paper, we have applied nonequilibrium Green functions simulations to the dynamics of small Hubbard clusters following strong excitations. In order to access larger times on the order of several 100 to 1000 inverse hopping amplitudes, we applied the generalized Kadanoff--Baym ansatz with Hartree--Fock propagators (HF--GKBA). This ansatz reduced the dynamics to a single-time dynamics while at the same time fully retaining conservation laws and memory effects. 
This was demonstrated explicitly using a density operator approach. In contrast, the use of the GKBA with propagators that contain correlation selfenergy contributions (non-hermitean selfenergy with a non-vanishing imaginary part) leads to a violation of the conservation laws, even if the original selfenergy in the two-time NEGF theory was conserving.

The question of total energy conservation in the GKBA was reconsidered in an NEGF framework in Sec.~\ref{ss:general_consreving}. It was shown that the theorem of Baym and Kadanoff does not directly apply to this approximation and we demonstrated how the condition of this theorem can be relaxed. As a result, a new class of conserving approximations where the retarded and the less Green function evolve with different selfenergies is introduced of which the GKBA is just one representative.

The HF-GKBA not only increases the computational efficiency of the calculations, it even shows a qualitatively improved behavior: artificial damping effects observed previously in two-time simulations of strongly driven Hubbard clusters [\onlinecite{friesen_10}] do not occur within the HF--GKBA. This was traced to the reduced degree of selfconsistency in the computation of the Green functions, compared to the full two-time case, cf. Fig.~\ref{fig:diagrams_sigma}. While in macroscopic systems 
this is a minor issue and single-time and two-time calculations show comparable accuracy [\onlinecite{bonitz_jpcm96,kwong98,freericks}], the present results indicate that,  for finite systems, reduction of the selfconsistency appears to be an important issue, in agreement with earlier analysis [\onlinecite{friesen_10,hermanns_jpcs13}].

Comparisons with exact diagonalization results for $N=4$ and $N=8$ allow us to conclude that the HF--GKBA is a reliable approach to the dynamics of weakly coupled ($U=0.1\dots 0.25$) Hubbard clusters during the initial relaxation period of $0\le t \lesssim 25\dots 40$. An exception is the case $N=2$ where quantitative agreement is limited to $t\lesssim 2$ (which is not surprising since any continuum-type approximation exhibits the strongest inaccuracies for small $N$). The accuracy of the HF--GKBA systematically improves for increasing $N$. 

We, therefore, expect that our HF--GKBA simulations also provide reliable results for $N\ge 16$. This will also allow to study 2D or 3D systems since the computational effort of the HF--GKBA depends only on the basis size, but is independent of the particle number and dimensionality.
\begin{figure}
    \includegraphics[width=9cm]{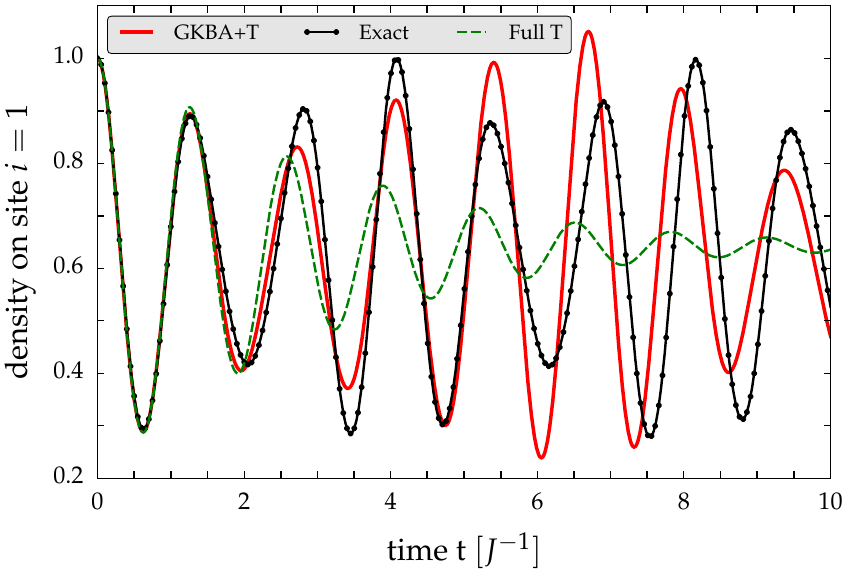}
\caption{(Color online) Same as Fig.~\ref{fig:verdozzi-vergleich}~(Top), $N=2$, but using T-matrix instead of 2B selfenergies.
The exact result (full black line) is compared to a full two-time T-matrix calculation (this result exactly matches the one of von Friesen et al. [\onlinecite{puigvonfriesen09}]) and the T-matrix with the HF-GKBA [\onlinecite{schluenzen14}].
}
\label{fig:verdozzi-vergleich-t}
\end{figure}

From the relaxation dynamics  we can draw the following conclusions. For the studied inhomogeneous initial state, there are four distinct stages:
\begin{description}
 \item[I.] Build-up of kinetic energy and decay of $n_1$ to $0.5$. This phase is a consequence of the particular inhomogeneous initial state and corresponds to a ballistic motion of the particles into the previously empty part of the cluster. This phase appears to be independent of the coupling strength $U$, but increases almost proportional with $N$.
 \item[II.] Build-up of correlation energy and saturation of Hartree--Fock energy. This phase becomes shorter when $U$ increases, but it extends with $N$.
 \item[III.] Relaxation of the site occupations to the homogeneous values of $0.5$. This scale grows with $U$ and is independent of $N$.
 \item[IV.] The fourth stage is characterized by revivals of the occupations. With increasing $U$ and $N$ these oscillation are more weakly pronounced.
\end{description}

It is interesting to compare these dynamics with earlier investigations. In fact, the short-time dynamics of interacting many-body and few-body systems have been analyzed for a variety of systems before. For the homogeneous electron gas or a dense plasma, studies of interaction quenches were performed in Refs.~[\onlinecite{bonitz_jpcm96}, \onlinecite{bonitz_pla96}, \onlinecite{bonitz_pre97}]. Here, the observation was that there exist two distinct phases: the first is characterized by the build up of correlations---manifest by the increase of (minus) the correlation energy and kinetic energy, i.e., by heating of the system. This effect has been termed correlation induced heating [\onlinecite{bonitz_pre97}] or disorder induced heating [\onlinecite{murillo}].
The duration of this stage is given by the correlation time $\tau_{\rm cor}$ which, in a homogeneous charged particle system, is of the order of the inverse of the plasma frequency. The second stage is characterized by a relaxation of the single-particle distribution function (occupations) and has a duration $t_{\rm rel}$--the relaxation time. Furthermore, negative quenches have been studied, where the interaction strength is rapidly reduced [\onlinecite{bonitz_pre97, semkat_pre99, semkat_jmp00,gericke_jpa03, gericke_jpa03_2}].
In this case, kinetic energy is reduced suggesting that the system can be cooled after it has been prepared in an ``overcorrelated'' initial state. Finally, we note that in various studies of moderately correlated macroscopic systems another kind of two-stage dynamics was observed where, preceding the final thermalization, a ``pre-thermalization'' plateau was found, e.g. [\onlinecite{berges_prl04,eckstein_prl09, eckstein_prb10}]. 

It remains an interesting question for future studies to analyze how these different scenarios are related to each other and how they depend on the coupling strength, system size and dimensionality. NEGF simulations connected with the HF-GKBA appear to be a powerful approach to this problem. This requires to  extend these simulations into the strong coupling range by using T-matrix selfenergies. A first result using T-matrix selfenergies, in combination with the HF-GKBA, is shown in 
Fig.~\ref{fig:verdozzi-vergleich-t} which is the same setup as Fig.~\ref{fig:verdozzi-vergleich}.a (except for the choice of the selfenergies). The two-time calculations in T-matrix approximation exhibit again artificial damping, as in the case of the second Born approximation. Again this damping is removed when the HF-GKBA is applied, and the overall accuracy increases. This gives further support for the concept of single-time NEGF calculations in the framework of the HF-GKBA. 
Details of the approximation and on its implementation are work in progress and will be reported elsewhere \cite{schluenzen14}.
\begin{acknowledgments}
MB acknowledges the hospitality of the University of Florida, Gainesville, where part of this work was performed.
This work is supported by the Deutsche Forschungsgemeinschaft via project BO1366-9 and by grant SHP006 for computer time at the HLRN.
\end{acknowledgments}

%\bibliography{qbm}

\begin{thebibliography}{10}%
\makeatletter
\providecommand \@ifxundefined [1]{%
 \ifx #1\undefined \expandafter \@firstoftwo
 \else \expandafter \@secondoftwo
\fi
}%
\providecommand \@ifnum [1]{%
 \ifnum #1\expandafter \@firstoftwo
 \else \expandafter \@secondoftwo
\fi
}%
\providecommand \enquote [1]{``#1''}%
\providecommand \bibnamefont  [1]{#1}%
\providecommand \bibfnamefont [1]{#1}%
\providecommand \citenamefont [1]{#1}%
\providecommand\href[0]{\@sanitize\@href}%
\providecommand\@href[1]{\endgroup\@@startlink{#1}\endgroup\@@href}%
\providecommand\@@href[1]{#1\@@endlink}%
\providecommand \@sanitize [0]{\begingroup\catcode`\&12\catcode`\#12\relax}%
\@ifxundefined \pdfoutput {\@firstoftwo}{%
 \@ifnum{\z@=\pdfoutput}{\@firstoftwo}{\@secondoftwo}%
}{%
 \providecommand\@@startlink[1]{\leavevmode}%
 \providecommand\@@endlink[0]{}%
}{%
 \providecommand\@@startlink[1]{%
  \leavevmode
  \pdfstartlink
   attr{/Border[0 0 1 ]/H/I/C[0 1 1]}%
   user{/Subtype/Link/A<</Type/Action/S/URI/URI(#1)>>}%
  \relax
 }%
 \providecommand\@@endlink[0]{\pdfendlink}%
}%
\providecommand \url  [0]{\begingroup\@sanitize \@url }%
\providecommand \@url [1]{\endgroup\@href {#1}{\urlprefix}}%
\providecommand \urlprefix [0]{URL }%
\providecommand \Eprint[0]{\href }%
\@ifxundefined \urlstyle {%
  \providecommand \doi [1]{doi:\discretionary{}{}{}#1}%
}{%
  \providecommand \doi [0]{doi:\discretionary{}{}{}\begingroup
  \urlstyle{rm}\Url }%
}%
\providecommand \doibase [0]{http://dx.doi.org/}%
\providecommand \Doi[1]{\href{\doibase#1}}%
\providecommand \bibAnnote [3]{%
  \BibitemShut{#1}%
  \begin{quotation}\noindent
    \textsc{Key:}\ #2\\\textsc{Annotation:}\ #3%
  \end{quotation}%
}%
\providecommand \bibAnnoteFile [2]{%
  \IfFileExists{#2}{\bibAnnote {#1} {#2} {\input{#2}}}{}%
}%
\providecommand \typeout [0]{\immediate \write \m@ne }%
\providecommand \selectlanguage [0]{\@gobble}%
\providecommand \bibinfo [0]{\@secondoftwo}%
\providecommand \bibfield [0]{\@secondoftwo}%
\providecommand \translation [1]{[#1]}%
\providecommand \BibitemOpen[0]{}%
\providecommand \bibitemStop [0]{}%
\providecommand \bibitemNoStop [0]{.\EOS\space}%
\providecommand \EOS [0]{\spacefactor3000\relax}%
\providecommand \BibitemShut [1]{\csname bibitem#1\endcsname}%
%</preamble>
\bibitem{pavarini11}{E.~Pavarini, E.~Koch, D.~Vollhardt and A.~Lichtenstein (Eds.), \textit{The LDA+DMFT approach to strongly correlated materials}, (Forschungszentrum J\"ulich GmbH, Zentralbibliothek, Verlag 2011).}

\bibitem{balzer_jpcs13} K. Balzer, S. Hermanns, and M. Bonitz, J. Phys. Conf. Ser. {\bf 427}, 012006 (2013)

\bibitem{becker12}{W.~Becker, X.J.~Liu, P.J.~Ho and H.J.~Eberly, 
%\textit{Theories of photoelectron correlation in laser-driven multiple atomic ionization}, 
Rev.~Mod.~Phys.~\textbf{84}, 1011 (2012).}

\bibitem{schuette_prl12} B. Sch\"utte, S. Bauch, U. Fr\"uhling, M. Wieland, M. Gensch, E. Pl\"onjes, T. Gaumnitz, A. Azima, M. Bonitz, and M. Drescher,
%Evidence for chirped Auger electron emission
Phys. Rev. Lett. {\bf 108}, 253003 (2012); S. Bauch, and M. Bonitz, Phys. Rev. A {\bf 85}, 053416 (2012)

\bibitem{kehrein_njp10} M. Moeckel, and S. Kehrein, New J. Phys. {\bf 12}, 055016 (2010).
\bibitem{eckstein_prl09} M. Eckstein, M. Kollar, and Ph. Werner, Phys. Rev. Lett. {\bf 103}, 056403 (2009).
\bibitem{eckstein_prb10} M. Eckstein, M. Kollar, and Ph. Werner, Phys. Rev. B. {\bf 81}, 115131 (2010).

\bibitem{bloch08}{I.~Bloch, J.~Dalibard and W.~Zwerger, 
%\textit{Many-body physics with ultracold gases}, 
Rev.~Mod.~Phys.~\textbf{80}, 885 (2008).}

\bibitem{thorwart_13} S. Weiss, R. H\"utzen, D. Becker, J. Eckel, R. Egger, M. Thorwart, arXiv:1304.6919
%Iterative path integral summation for nonequilibrium quantum transport

\bibitem{book_bonitz_qkt}{M.~Bonitz: \textit{Quantum Kinetic Theory} (Teubner, Stuttgart, Leipzig, 1998).}

\bibitem{akbari_12} A. Akbari, M.J. Hashemi, A. Rubio, R.M. Nieminen, and R. van Leeuwen, Phys. Rev. B {\bf 85}, 235121 (2012).

\bibitem{hermanns_jpcs13} S. Hermanns, K. Balzer, and M. Bonitz,
%Few-particle quantum dynamics - comparing Nonequilibrium Green functions with the generalized Kadanoff--Baym ansatz to density operator theory
J. Phys. Conf. Ser. {\bf 427}, 012008 (2013).

\bibitem{ayik09}
S. Ayik, Phys. Lett. B {\bf 658}, 174 (2008)
%A stochastic mean-field approach for nuclear dynamics

\bibitem{lacroix14} D. Lacroix, S. Hermanns, C. Hinz, and M. Bonitz, submitted for publication, arXiv:1403.5098

\bibitem{kwong98}{N.H.~Kwong, M.~Bonitz, R.~Binder and H.S.~K\"ohler, 
%\textit{Semiconductor Kadanoff--Baym equation results for optically excited electron-hole plasmas in quantum wells}, 
phys.~stat.~sol.~(b)~\textbf{206}, 197 (1998).}

\bibitem{kwong00}{N.-H.~Kwong and M.~Bonitz, 
%\textit{Real-time Kadanoff--Baym approach to plasma oscillations in a correlated electron gas}, 
Phys.~Rev.~Lett.~\textbf{84}, 1768 (2000).}

\bibitem{rios11}{A.~Rios, B.~Barker, M.~Buchler and P.~Danielewicz, 
%\textit{Towards a nonequilibrium Green function description of nuclear reactions: One-dimensional mean-field dynamics}, 
Annals of Physics~\textbf{326}, 1274 (2011).}

\bibitem{bonitz_cpp99} M. Bonitz, Th. Bornath, D. Kremp, M. Schlanges, and W.D. Kraeft,
%Quantum kinetic theory for laser plasmas. Dynamical screening in strong fields,
Contrib. Plasma Phys. {\bf 39}, 329 (1999).

\bibitem{haberland_pre01} H. Haberland, M. Bonitz, and D. Kremp,
%Harmonics generation in electron-ion collisions in a strong laser pulse
Phys. Rev. E {\bf 64}, 026405 (2001).

\bibitem{garny11}{M.~Garny, A.~Kartavtsev and A.~Hohenegger, 
%\textit{Leptogenesis from first principles in the resonant regime}, arXiv:1112.6428.
Annals of Physics {\bf 328}, 26 (2013)}.

\bibitem{bornath99} T. Bornath, D. Kremp, and M. Schlanges Phys. Rev. E {\bf 60}, 6382 (1999)

\bibitem{bonitz_jpcs13} M. Bonitz, S. Hermanns, K. Kobusch, and K. Balzer,
J. Phys. Conf. Ser. {\bf 427}, 012002 (2013)

\bibitem{bonitz_cpp99_2} M. Bonitz, N.H. Kwong, D. Semkat, and D. Kremp,
%Generalized Kadanoff-Baym theory for nonequilibrium many-body systems in external fields. An effective multi-band approach,
Contr. Plasma Phys. {\bf 39}, 37 (1999) 

\bibitem{gartner06}{P.~Gartner, J.~Seebeck and F.~Jahnke, 
%\textit{Relaxation properties of the quantum kinetics of carrier-LO-phonon interaction in quantum wells and quantum dots}, 
Phys.~Rev.~B~\textbf{73}, 115307 (2006).}

\bibitem{lorke06}{M.~Lorke, T.R.~Nielsen, J.~Seebeck, P.~Gartner and F.~Jahnke, 
%\textit{Influence of carrier-carrier and carrier-phonon correlations on optical absorption and gain in quantum-dot systems}, 
Phys.~Rev.~B~\textbf{73}, 085324 (2006).}

\bibitem{bonitz_prb07} M. Bonitz, K. Balzer, and R. van Leeuwen,
%Invariance of the Kohn (sloshing) mode in a conserving theory,
%ArXiv: cond-mat/0702633
Phys. Rev. B {\bf 76}, 045341 (2007). 

\bibitem{balzer_prb09} K. Balzer, M. Bonitz, R. van Leeuwen, N.E. Dahlen, and A. Stan,
%Nonequilibrium Green functions approach to strongly correlated few-electron quantum dots
Phys. Rev. B {\bf 79},  245306 (2009).

\bibitem{uimonen11} A.-M.~Uimonen, E.~Khosravi, A.~Stan, G.~Stefanucci, S.~Kurth and R.~van Leeuwen and E.K.U.~Gross, 
%\textit{Comparative study of many-body perturbation theory and time-dependent density functional theory in the out-of-equilibrium Anderson model},
Phys.~Rev.~B~\textbf{84}, 115103 (2011).

\bibitem{khosravi12}{E.~Khosravi, A.-M.~Uimonen, A.~Stan, G.~Stefanucci, S.~Kurth, R.~van Leeuwen and E.K.U.~Gross, 
%\textit{Correlation effects in bistability at the nanoscale: Steady state and beyond}, 
Phys.~Rev.~B~\textbf{85}, 075103 (2012).}

\bibitem{dahlen07} N.E.~Dahlen, and R. van Leeuwen, Phys. Rev. Lett. {\bf 98}, 153004 (2007). 

\bibitem{balzer_pra10} K. Balzer, S. Bauch, and M. Bonitz
%Efficient grid-based method in nonequilibrium Green’s function calculations: Application to model atoms and molecules
Phys. Rev. A {\bf 81}, 022510 (2010). 

\bibitem{balzer_pra10_2} K. Balzer, S. Bauch, and M. Bonitz
%Time-dependent second-order Born calculations for model atoms and molecules in strong laser fields
Phys. Rev. A {\bf 82}, 033427 (2010). 

\bibitem{balzer_lnp13} K. Balzer, and M. Bonitz, {\em Nonequilibrium Green Functions Approach to Inhomogeneous Systems}, 
Lecture Notes in Physics, vol. 867, Springer (2013).

\bibitem{stefanucci_book_13} G. Stefanucci, and R. van Leeuwen, 
{\em Nonequilibrium Many-Body Theory of Quantum Systems: A Modern Introduction }, Oxford, 2013 

\bibitem{puigvonfriesen09}{M.~Puig von Friesen, C.~Verdozzi and C.-O.~Almbladh,
%\textit{Successes and failures of Kadanoff--Baym dynamics in Hubbard nanoclusters}, 
Phys.~Rev.~Lett.~\textbf{103}, 176404 (2009).}

\bibitem{puigvonfriesen10}{M.~Puig von Friesen, C.~Verdozzi and C.-O.~Almbladh, 
%\textit{Kadanoff--Baym dynamics of Hubbard clusters: Performance of many-body schemes, correlation-induced damping and multiple steady and quasi-steady states}, 
Phys.~Rev.~B~\textbf{82}, 155108 (2010).}

\bibitem{balzer12_epl}{K.~Balzer, S.~Hermanns and M.~Bonitz,
%\textit{Electronic double-excitations in quantum wells: Solving the two-time Kadanoff--Baym equations}, 
EPL~\textbf{98}, 67002 (2012).}

\bibitem{sakkinen12}{N.~S\"akkinen, M.~Manninen and R.~van Leeuwen,
%\textit{The Kadanoff--Baym approach to double excitations in finite systems}, 
New J.~Phys.~\textbf{14}, 013032 (2012).}

\bibitem{lipavsky86} P.~Lipavsk{\'y}, V.~\ifmmode \check{S}\else \v{S}\fi{}pi\ifmmode \check{c}\else \v{c}\fi{}ka and B.~Velick{\'y}, 
%\textit{Generalized Kadanoff--Baym ansatz for deriving quantum transport equations}, 
Phys.~Rev.~B~\textbf{34}, 6933 (1986).

\bibitem{hermanns12_pysscripta} S.~Hermanns, K.~Balzer and M.~Bonitz, 
%\textit{Nonequilibrium Green functions approach to inhomogeneous quantum many-body systems using the generalized Kadanoff--Baym ansatz}, 
Physica~Scripta {\bf T151}, 014035 (2012).

\bibitem{bonitz_cpp13} M. Bonitz, S. Hermanns, and K. Balzer, Contrib. Plasma Phys. {\bf 53}, 778 (2013), arXiv:1309.4574

\bibitem{stefanucci_gkba} S. Latini, E. Perfetto, A.-M. Uimonen, R. van Leeuwen, and G. Stefanucci, Phys. Rev. B {\bf 89}, 0753006 (2014)

\bibitem{lev_14} Y. Bar Lev, and D.R. Reichman, arXiv:1402.0502

\bibitem{KeldyshContour} L.V. Keldysh, ZhETF \b{47}, 1515 (1964).

\bibitem{martin_theory_1959} P. Martin, and J. Schwinger, Phys, Rev, {\bf 115}, 1342 (1959).

\bibitem{book_kadanoffbaym_qsm}{L.P.~Kadanoff and G.~Baym: \textit{Quantum Statistical Mechanics} (Benjamin, New York, 1962).}

\bibitem{baym61} G. Baym and L.P. Kadanoff, Phys. Rev. {\bf 124}, 287 (1961).
%Conservation laws and correlation functions

\bibitem{book_bonitz_semkat} \textit{Introduction to Computational Methods in Many-Body Physics},
M. Bonitz, and D. Semkat (Eds.), Rinton Press, Princeton, 2006. 

\bibitem{semkat_jmp00} D. Semkat, D. Kremp, and M. Bonitz, 
%\textit {Kadanoff--Baym equations and non-Markovian Boltzmann equation in generalized T-matrix approximation}
J. Math. Phys. {\bf 41}, 7458 (2000).

\bibitem{stefanucci13} R van Leeuwen and G Stefanucci,
%\textit{Equilibrium and nonequilibrium many-body perturbation theory: a unified framework based on the Martin-Schwinger hierarchy} 
J. Phys. Conf. Ser. {\bf 427}, 012001 (2013).

\bibitem{fedvr} For continuous systems substantial advances have recently been achieved via the choice of special basis representations (FEDVR basis), e.g., \cite{balzer_pra10}, for lattice systems this problem does not occur.

\bibitem{LWR}
David~C. Langreth and John~W. Wilkins.
%\newblock Theory of spin resonance in dilute magnetic alloys.
Phys. Rev. B, {\bf 6}, 3189 (1972).

\bibitem{bonitz_jpcm96} M. Bonitz, D. Kremp, D.C. Scott, R. Binder, W. D. Kraeft, and H. S. K\"ohler,
%Numerical analysis of memory effects in the intraband relaxation in semiconductors
Journal of Physics: Condensed Matter {\bf 8}, 6057  (1996)

\bibitem{bonitz_pla96} M. Bonitz, D. Kremp,
%Kinetic energy relaxation and correlation time of nonequilibrium many-particle systems
Phys. Lett. A {\bf 212}, 83 (1996) 

\bibitem{bonitz_epjb99} M. Bonitz, D. Semkat and H. Haug,
%Non-Lorentzian spectral functions for Coulomb quantum kinetics,
Europ. Phys. J. B {\bf 9}, 309 (1999) 

\bibitem{kremp_ap97} D. Kremp, M. Bonitz, W.D. Kraeft, and M. Schlanges, Ann. Phys. (N.Y.) E {\bf 258}, 320 (1997)
\bibitem{dufty_boercker} D.B. Boercker, and J.W. Dufty, Ann. Phys. (N.Y.) {\bf 119}, 43 (1979). 
\bibitem{friesen_10}{M.~Puig von Friesen, C.~Verdozzi and C.-O.~Almbladh,
%\textit{Kadanoff--Baym equations and approximate double occupancy in a Hubbard dimer}, 
arXiv:1009.2917 (2010).}
\bibitem{freericks} Turkowski and Freericks (Phys. Rev. B) showed that, for macroscopic systems the GKBA does not reproduce all sum rules of the spectral function.
\bibitem{bonitz_pre97} M. Bonitz, D. Semkat, and D. Kremp, Phys. Rev. E {\bf 56}, 1246 (1997).
\bibitem{murillo} M.S. Murillo, Phys. Rev. Lett. {\bf 87}, 115031 (2001).
\bibitem{semkat_pre99} D. Semkat, D. Kremp, and M. Bonitz, Phys. Rev. E {\bf 59}, 1557 (1999).
%\bibitem{morawetz_pre01} K. Morawetz, M. Bonitz, V.G. Morozov, G. R\"opke, and D. Kremp, Phys. Rev. E {\bf 63}, 020102 (2001).
\bibitem{gericke_jpa03} D.O. Gericke, M.S. Murillo, D. Semkat, M. Bonitz, and D. Kremp, J. Phys. A: Math. Gen. {\bf 36}, 6087 (2003).
%Relaxation of Strongly Coupled Coulomb Systems after Rapid Changes of the interaction potential 
\bibitem{gericke_jpa03_2} D.O. Gericke, M.S. Murillo,  M. Bonitz, and D. Semkat, J. Phys. A: Math. Gen. {\bf 36}, 6095 (2003).
%Temperature estimates for quantum systems after an ionization iduced rapid switch of the spin statistics 
\bibitem{berges_prl04} J. Berges, Sz. Borsanyi, and C. Wetterich, Phys. Rev. Lett. {\bf 93}, 142002 (2004).

%new refs
% energy conservation with density operators
\bibitem{dufty_boercker89} J.W. Dufty and D.B. Boercker, J. Stat. Phys. {\bf 57}, 827 (1989). 

\bibitem{dufty_97} J.W. Dufty, Contrib. Plasma Phys. {\bf 37}, 129 (1997). 

\bibitem{schluenzen14} N. Schl\"unzen, S. Hermanns, and M. Bonitz, to be published

\end{thebibliography}

%Merlin.mbs v4.21 2009-07-09.
\providecommand{\noopsort}[1]{}\providecommand{\singleletter}[1]{#1}%

\end{document}